\documentclass[12pt]{article}
\usepackage{amssymb, enumitem, amsmath, booktabs}
\usepackage{graphicx}
\usepackage{enumerate}
\usepackage{xcolor}
\usepackage{subcaption}
\usepackage{tabularx}
\usepackage{nccmath}
\usepackage{threeparttable, tablefootnote}
\usepackage{natbib}
\usepackage{url} 
\newlist{steps}{enumerate}{1}
\setlist[steps, 1]{label = Step \arabic*:}
\newcommand{\blind}{1}

\date{}
\begin{document}

\def\spacingset#1{\renewcommand{\baselinestretch}%
{#1}\small\normalsize} \spacingset{1}


\if1\blind
{
  \title{\bf Learning associations of COVID-19 hospitalizations with wastewater viral signals by Markov modulated models}
  \author{K. Ken Peng\\
    Department of Statistics and Actuarial Science, Simon Fraser University\\
    Charmaine B. Dean \\
    Department of Statistics and Actuarial Science, University of Waterloo\\
    Robert Delatolla \\
    Department of Civil Engineering, University of Ottawa\\
    X. Joan Hu \\
    Department of Statistics and Actuarial Science, Simon Fraser University\\
   }
  \maketitle
} \fi

\if0\blind
{
  \bigskip
  \bigskip
  \bigskip
  \begin{center}
    {\LARGE\bf Exploring associations of COVID-19 hospitalizations with wastewater viral signals by Markov modulated model with distributed lag}
\end{center}
  \medskip
} \fi

\bigskip
\begin{abstract}
Viral signal in wastewater offers a promising opportunity to assess and predict the burden of infectious diseases. That has driven the widespread adoption and development of wastewater monitoring tools by public health organizations. Recent research highlights a strong correlation between COVID-19 hospitalizations and wastewater viral signals, and validates that increases in wastewater measurements may offer early warnings of an increase in hospital admissions. Previous studies \citep[e.g.][]{peng2023exploration} utilize distributed lag models to explore associations of COVID-19 hospitalizations with lagged SARS-CoV-2 wastewater viral signals. However, the conventional distributed lag models assume the duration time of the lag to be fixed, which is not always plausible. This paper presents Markov-modulated models with distributed lasting time, treating the duration of the lag as a random variable defined by a hidden Markov chain. We evaluate exposure effects over the duration time and estimate the distribution of the lasting time using the wastewater data and COVID-19 hospitalization records from Ottawa, Canada during June 2020 to November 2022. The different COVID-19 waves are accommodated in the statistical learning. In particular, two strategies for comparing the associations over different time intervals are exemplified using the Ottawa data. Of note, the proposed Markov modulated models, an extension of distributed lag models, are potentially applicable to many different problems where the lag time is not fixed.
\end{abstract}

\noindent%
{\it Keywords:} Distributed lag models, hidden Markov process, stratified analysis wastewater surveillance
\vfill

\newpage
\spacingset{1.9} 
\section{Introduction}

\subsection{Background}

During infection with the SARs-CoV-2 virus, for example, SARS-CoV-2 ribonucleic acid (RNA) was excreted into the sewer system via feces, saliva, swabs, and/or sputum of infected individuals \citep{anand2021review, peccia2020measurement}. That wastewater viral signal has been successfully detected in wastewater surveillance systems using the \textit{RT-PCR technique} \citep{medema2020presence, randazzo2020sars}. When testing COVID-19 infection is limited, the wastewater viral signal has been taken as an effective biological marker to track the prevalence of COVID-19. Instead of a direct relation between the wastewater viral signals and COVID-19 epidemiological metrics at a specific time point, a time lag was found to exist between the appearance of increased SARS-CoV-2 RNA signal in wastewater and certain epidemiological metrics \citep{larsen2020tracking}. Many recent studies use the distributed lag models (DLM) to explore the association between wastewater viral signal and COVID-19 epidemiology \citep{galani2022sars, peccia2020measurement, kaplan2021aligning, zulli2022predicting, schoen2022sars, xie2022rna, peng2023exploration}. More broadly, DLMs have been also widely employed in social sciences \citep{judge1982introduction} and epidemiology for other diseases \citep{pope1996time}.

When learning how one quantity  is associated with an exposure by DLMs, often is of interest to investigate the critical window of the exposure, that is, how far the exposure in the past significantly affects the current response. In general, specifying a correct maximal time lag (or lasting time), beyond which there is no exposure effect, is challenging. Inconsistent estimates may occur due to the lag is specified to be either shorter or longer than its actual value \citep{dhrymes1971strong}, and thus it yields failing to determine the critical window of exposure effects. Different ways to choose the lasting time have been proposed previously. First, one may consider it as a variable selection procedure and truncate the lag to a finite length. For example, \cite{davidson1993estimation} suggests to start with a reasonable maximum time lag, and then 
check whether the fit of a predetermined model deteriorates as the maximum time lag is reduced based on some criterion such as AIC and BIC. \cite{tibshirani2016ordered} consider an order-constrained lasso under the monotone constraint of the regression coefficients to the exposures overtime, which shrinks the delayed effect to approach 0 for longer time lags. Other approaches specify the exposure effects overtime into some functional forms that allow the effects to gradually decay to zero. For example, \cite{koyck1954distributed} proposed the geometric lag, \cite{almon1965distributed} proposed the polynomial lag, \cite{jorgenson1966rational} proposed a rational lag, \cite{tsurumi1971note} proposed the gamma lag, and \cite{ghysels2007midas} proposed using the exponential Almon lag. In the continuous-time case, one may consider historical functional regression \citep{malfait2003historical}.

\begin{figure}[!htp]
     \centering
		\includegraphics[width=1\linewidth]{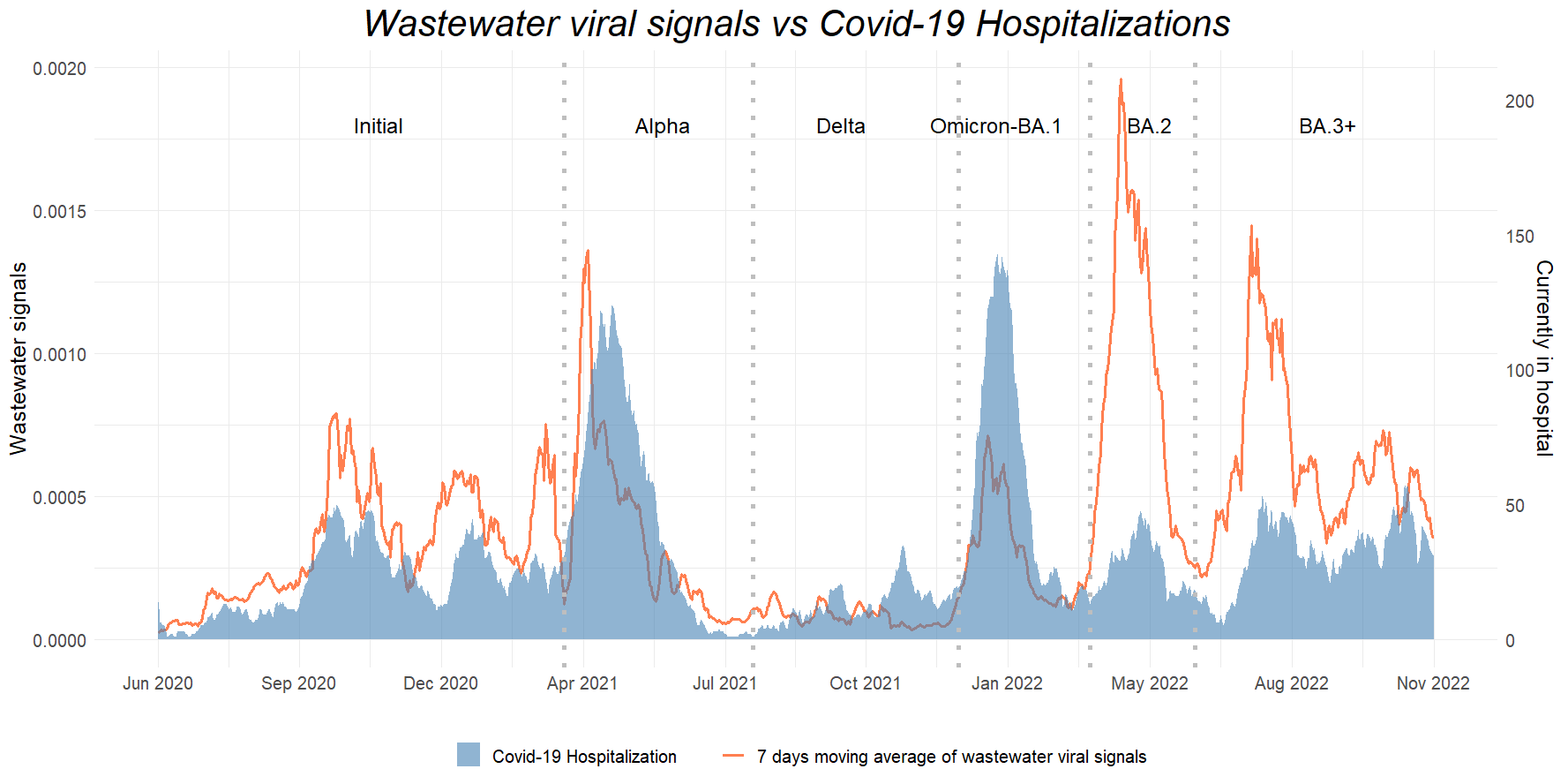}
		\caption{This figure illustrates the delayed association between Wastewater viral signals \& hospitalizations in Ottawa from 2020-06-16 to 2022 Nov 13rd. The blue area shows the number of patients currently in hospitals caused by COVID-19, and the orange curve shows the 7-day moving average of wastewater viral signal.}
  \label{rawdata}
		\end{figure}

Ottawa is the capital of Canada, a city with the population of approximate one million inhabitants. Wastewater samples at Ottawa are collected 7 days a week, analyzed, posted on GitHub and reported to Ottawa Public Health. Detailed methodology of detecting and quantifying the wastewater surveillance processes are further described in \cite{d2021catching}. We focus on the study period from June 16th, 2020, to November 13rd, 2022. The SARS-CoV-2 measurements after November 13rd, 2022 are related to variants of COVID-19 in Ottawa and require further model considerations. There are 60 missing dates which occur mostly between June and September 2020, as the daily collection started on September 2020. We imputed those 60 missing values of viral load using a smoothing spline regression. Figure \ref{rawdata} overlays the hospitalizations due to COVID-19 and the 7-day moving average of the wastewater viral signals. We observe that the trend of wastewater viral signals precedes the trend of COVID-19 hospitalizations, and there is a delay in the COVID-19 hospitalizations increase after the wastewater viral signals increase. That delay can be interpreted as the average gap time between an individual being infected and the individual being hospitalized. \cite{peng2023exploration} performed a correlation analysis to investigate how the delay varies over time. 

Upon the knowledge of the authors, all distributed lag models in the literature to date consider the maximum time lag as fixed. However, at maximum time lag likely varies across subjects or time. For instance, the delay between the increase of viral signals in wastewater and COVID-19 hospitalizations may change depending on the dominant variant or the specific region. In this paper, we propose a Markov-modulated model with distributed lasting time, framing the maximum time lag (lasting time) in the conventional DLMs as a random variable defined by a hidden stochastic process. Our model allows flexible shapes of the probability distribution function of the lasting time. And we explore the associations of risk to COVID-19 hospitalizations with wastewater viral signals using the data from Ottawa, Canada and evaluate the exposure effects with the proposed model. 

The manuscript is organized as follows: In Section 2, the proposed model is presented and exemplified by its
several useful special cases. Section 3 shows estimation procedures for the model parameters and analyses of the aforementioned Ottawa wastewater data. We provide some final remarks in Section 4.

\section{Notation and modeling}

\subsection{Distributed lag models with random lasting times}

Consider two series of measurements on response $Y$ and exposure $X$, denoted by $Y_t$ and $X_t$ with times $t = 1,...,T$, respectively. Let $\mathcal{H}_x(t) = \{X_t, X_{t-1},...,X_0\}$ be all the $X$ history information upto time $t$. We introduce $\tau = 0,...,t$ for the backwards time (or time lag) starting from $t$. The lasting time of the delayed effect is defined as the largest $\tau$ that $Y_t$ is associated with $X_{t-\tau}$ for a given $t$. In engineering literature, the term ``lead time" is often used to describe the optimal time-lag between, say, wastewater viral signals (the exposure) and COVID-19 hospitalizations (the response outcome) to maximize their correlations, e.g. \cite{hegazy2022understanding}. 

To investigate how the association between $Y_t$ and $\mathcal{H}_x(t)$ in the presence of other
covariates $\pmb{W}$, we propose an extension of the conventional distributed lag models, named Markov modulated models with distributed lasting time:
\begin{align}
\begin{split}
    & E\big(Y_t\big|\mathcal{H}_x(t), \pmb{W}, \{Z^*_t(\tau), 0 \leq \tau \leq t\}; \alpha_0, \pmb{\alpha}, \pmb{\beta}^*\big) 
    = \alpha_0 + \pmb{W}'\pmb{\alpha} + \sum_{\tau=0}^t\beta_{\tau}(t)X_{t-\tau} \\
    & \text{ with } 
    \beta_{\tau}(t) =  \beta^*_{\tau}(t)Z^*_t(\tau), 
\label{dlm-markov}
\end{split}
\end{align}
where $\alpha_0$ is the intercept, $\pmb{\alpha}$ is the regression coefficients to $\pmb{W}$, 
$\{Z^*_t(\tau), 0 \leq \tau \leq t\}$ is a segment of a stochastic process with states $0$ and $1$,
and $\beta^*_{\tau}(t)$ is the regression coefficient associated with the primary exposure at the time $t-\tau$ when $Z^*_t(\tau) = 1$. This model includes models considered in, for example, \cite{dhrymes1971strong} and \cite{malfait2003historical} as special cases. Note that the regression coefficient $\beta_{\tau}(t) = \beta^*_{\tau}(t)Z^*_t(\tau)$ is random in general since $Z^*_t(\tau)$ is random. 

The process $\{Z^*_t(\tau), 0 \leq \tau \leq t \}$ is unobservable (i.e. hidden). 
Assume that $\{Z^*_t(\tau), 0 \leq \tau \leq t \}$ starts from the state $1$ with probability $1$, i.e. $P(Z^*_t(0) = 1) \equiv 1$, for all $t$. This assumes $Y_t$ is always associated to the explanatory variable $X_t$. Further assume that $\{Z^*_t(\tau), 0 \leq \tau \leq t \}$ transitions from state $1$ to state $0$ at each time point with
certain probability, but not back from state $0$ to state $1$. That is, if $Y_t$ is not associated with the $X_{t-\tau}$, it is not associated with the exposure $X$ from further back. Thus the process $\{Z^*_t(\tau), 0 \leq \tau \leq t \}$ is a hidden semi-Markov chain with the transition probability matrix
 \begin{align}
P(\tau,t) = \begin{bmatrix}
    1 & 0  \\
    \rho(\tau,t) & 1-\rho(\tau,t) \\
\end{bmatrix},
\end{align} 
where $0 < \rho(\tau,t) < 1$ is the transition probability $P(Z^*_t(\tau+1) = 0|Z^*_t(\tau) = 1)$. The transition probability can be further specified as fixed, a function of $\tau$ and $t$, or depends on potential covariates $\pmb{W}$. Under the model in (\ref{dlm-markov}), the lasting time can be written as $L(t) = max\{\tau: Z^*_t(\tau) = 1, 0\leq \tau\leq t\}$, a discrete random variable whose distribution depends on $\rho(\tau, t)$. The hidden process $\{Z^*_t(\tau), 
0 \leq \tau \leq t\}$ yields different realizations of $L(t)$ with a given $t$, denoted by $l_t$, which is also unobservable.

\subsection{Distribution of the lasting time}

The lasting time $L(t) = max\{\tau: Z^*_t(\tau) = 1, 0\leq \tau\leq t \}$ of given $t$. The probability mass function can be calculated as follows:
\begin{align*}
& P(L(t) = \tau) = P(Z^*_t(\tau+1)=0, Z^*_t(\tau)=1) \\
& = P(Z^*_t(\tau+1)=0|Z^*_t(\tau)=1)P_t(Z^*_t(\tau)=1|Z^*_t(\tau-1))...P_t(Z^*_t(1)=1|Z^*_t(0)=1) \\
& = \rho(\tau,t)\prod_{i=0}^{\tau-1}(1-\rho(i,t)).
\end{align*}
The cumulative probability function of $L(t)$ is thus $P(L(t) \leq \tau) = \sum^{\tau}_{i=0}P(L(t) = i)$.
The expectation of $Z^*_t(\tau)$, the random term of the parameter $\beta_{\tau}(t)$, is $E(Z^*_t(\tau)) = P(L(t) \geq \tau) = 1-\sum^{\tau-1}_{i=0}P(L(t) = i)$, and $E(\beta_{\tau}(t)) = \beta^*_{\tau}(t) E(Z^*_t(\tau))$.

\subsection{Specifications of model components}

The term $\beta_{\tau}(\cdot)$ in the proposed model (\ref{dlm-markov}) formulates the associations between the response $Y$ and the exposure $X$ consists of the ``conventional" and the ``stochastic" parts. The ``conventional" part $\beta^*_{\tau}(t)$ can be a smooth function of time lag $\tau$ for a given $t$, similar to the conventional DLMs. The ``stochastic" part $\{Z^*_t(\tau), 0 \leq \tau \leq t \}$ defines the lasting time $L(t)$. Our proposed model includes the conventional DLMs as special cases with the corresponding specifications of its ``conventional" and ``stochastic" components. 

For example, for any fixed $t$, with $\beta^*_{\tau}(t) \equiv \beta^*_{\tau}, \tau = 0,...,t$ , $\rho(\tau, t) = 1$ for $\tau \geq 31$, and $\rho(\tau, t) = 0$ for $\tau \leq 30$, model (\ref{dlm-markov}) reduces to distributed lag linear model (1) with $L(t) \equiv 30$. If further assume $\beta^*_{\tau}(t) = \beta^*_{\tau}$ to be a polynomial function of $\tau$, model (\ref{dlm-markov}) reduces to the popular Almon distributed lag model. 
In the following subsections, we present some important special cases that were motivated by our learning of the wastewater surveillance and COVID-19 hospitalizations.

\medskip

\noindent {\sl Example 1: $\{Z^*_t(\tau), 0 \leq \tau \leq t \}$ is a Markov chain for any fixed $t$.}

The simplest specification of model (\ref{dlm-markov}) is that $\beta^*_{\tau}(t)$ are unknown constants $\beta^*_0,...,\beta^*_t$ for any fixed $t$, and $\rho(\tau, t) \equiv \rho$ with $\rho$ an unknown parameter. In this case, $\{Z^*_t(\tau), 0 \leq \tau \leq t \}$ forms a two-state Markov chain with the transitional probability matrix
$ \begin{bmatrix}
    1 & 0  \\
    \rho & 1-\rho \\
\end{bmatrix}$. 
Our proposed model is then
\begin{align}
\begin{split}
       & E(Y_t|\mathcal{H}_x(t), \{Z^*_t(\tau), 0 \leq \tau \leq t \}; \alpha_0, \pmb{\beta}^*) 
    = \alpha_0 + \sum_{\tau=0}^t\beta_{\tau}X_{t-\tau} \\
    & \text{ with } \beta(\tau) = \beta_\tau^*Z^*_t(\tau).
\label{case1}
\end{split}
\end{align}

This model specification is suitable when the effect of $X$ on $Y$ exhibits a decay trend across time lag $\tau$. With a fixed $\rho>0$, both the probability mass function $P(L(t) = \tau|\rho)$ and $E(Z^*_t(\tau)|\rho)$ are monotonically decreasing. This characteristic shrinks $E(\beta_{\tau}|\rho)$ to $0$ for sufficiently large $\tau$. 

\medskip

\noindent {\sl Example 2: $\{Z^*_t(\tau), 0 \leq \tau \leq t \}$ is a semi-Markov chain.}

Still assuming $\beta^*_{\tau}(t)$ as unknown constants $\beta^*_0,...,\beta^*_t$ for any given $t$. With $\pmb{\lambda}$ a parameter vector $(\lambda_0,\lambda_1,\ldots, \lambda_J)'$, we consider the transition probability $\rho(\tau,t)$ of the ``stochastic" term $\{Z^*_t(\tau), 0 \leq \tau \leq t \}$ to be a step function of time $t$: $\rho(\tau, t;\pmb{\lambda}) = \frac{1}{1+e^{\lambda_0+\lambda_j\tau}}$ for $t \in (t_{j-1},t_j]$, $j=1,\ldots, J$, where $\{t_j: j =0, 1,...,J\}$ split the entire study period into $J$ sub-intervals with $t_0$ and $t_J$ to be the start and end dates of the study period. The hidden process $\{Z^*_t(\tau), 0 \leq \tau \leq t \}$ is a two-state semi-Markov chain. The proposed model is then
\begin{align}
\begin{split}
         &E(Y_t|\mathcal{H}_x(t), \{Z^*_t(\tau), 0 \leq \tau \leq t\}; \alpha_0, \pmb{\alpha}, \pmb{\beta}^*) 
    = \alpha_0+\pmb{\alpha}^{'}\pmb{W}
    + \sum_{\tau=0}^t\beta_{\tau}( t;\pmb{\lambda})X_{t-\tau} \\
    &\text{ with } 
    \beta_{\tau}(t;\pmb{\lambda}) = \beta^*_{\tau} Z^*_t(\tau)  
\label{case2}
\end{split}
\end{align}
where the components of $\pmb{W}$ are the period indicates $I\big(t\in(t_{j-1},t_j)\big)$ for $j=2,\ldots, J$, and the transitional probability matrix of $\{Z^*_t(\tau), 0 \leq \tau \leq t \}$ is 
\begin{equation*}
P(\tau, t;\pmb{\lambda}) = \begin{bmatrix}
    1 & 0  \\
    \rho(\tau, t;\pmb{\lambda}) & 1-\rho(\tau, t;\pmb{\lambda}) \\
\end{bmatrix} 
\end{equation*}

This model specification allows the probability mass function of $L(t)$ to be nonmonotone across $\tau$ and vary for
different time interval. Those features are desirable when there is lack of prior knowledge on the trend in the effect size of $X$ on $Y$ across different time lags $\tau$, or the associations between $Y$ and $X$ likely change with $t$ over different time periods. A reduced model of this example, which is an extension of the model in \textit{Example 1}, is provided in Section 1 of the Supplementary Material. It specifies $\{Z^*_t(\tau), 0 \leq \tau \leq t \}$ as a Markov chain for $t$ in each of the different time periods. 

\medskip

\noindent {\sl Example 3: ``Hard" stratified analysis for different time periods}

\textit{Example 2} assumes that different periods only differ in the distribution of the lasting time $L(t)$ but share the same set of the underlying $\beta^*_{\tau}, \tau = 0,...,t$. Since different time periods may differ in $\pmb{\beta}^*_{\tau}$, one may consider
the following model:
\begin{align}
\begin{split}
& E(Y_t|\mathcal{H}_x(t), \{Z^*_t(\tau), 0 \leq \tau \leq t\}; \pmb{\alpha}_0, \pmb{\beta}^{*}) = \sum^J_{j=1}I\big(t\in(t_{j-1},t_j]\big)\alpha^{(j)}_0 + \sum_{\tau=0}^t\beta_{\tau}(t;\pmb{\lambda})X_{t-\tau}\\
& \text{where } 
    \beta_{\tau}(t;\pmb{\lambda}) = \sum^J_{j=1}I\big(t\in(t_{j-1},t_j]\big)\beta^{*(j)}_{\tau} Z^{*}_t(\tau),
\label{case3}
\end{split}
\end{align}
and the transitional probability matrix of $\{Z^{*}_t(\tau), 0 \leq \tau \leq t \}$ is
$\begin{bmatrix}
    1 & 0  \\
    \rho(\tau,t|\pmb{\lambda}) & 1-\rho(\tau,t|\pmb{\lambda}) \\
\end{bmatrix}$
with $\rho(\tau,t|\pmb{\lambda}) = 
\frac{1}{1+e^{\lambda^{(j)}_0+\lambda^{(j)}_1\tau}}I\big(t\in(t_{j-1},t_j]\big)$.

An analysis under this model is essentially a stratified analysis for the different periods, using a separate model for each of the time periods. It is suitable when the cut-offs between periods are well-defined. We name the stratification with this model by a ``\textit{Hard}" stratification given the dates splitting the periods $\{t_j, j = 1,...,J\}$ are fixed and known.

\medskip

\noindent {\sl Example 4: ``Soft" stratified analysis for different time periods}

When the cut-offs between the periods are unclear, one may not be able to specify $\{t_j, j = 1,...,J\}$ as needed in the model  of Example 3. This situation is rather common in the infectious diseases processes. For example, the COVID-19 related hospitalizations at day $t$ may be caused by a mixture of more than one variant. To accommodate such situations, we propose a ``\textit{Soft}" stratification model as follows:
\begin{align}
\begin{split}
& E(Y_t|\mathcal{H}_x(t), \{Z^{*(j)}_t(\tau), 0 \leq \tau \leq t\}; \pmb{\alpha}_0, \pmb{\beta}^{*}) = \sum^J_{j=1}\pi_j(t)\alpha^{(j)}_0 + \sum_{\tau=0}^t\beta_{\tau}(t;\pmb{\lambda})X_{t-\tau},\\
& \text{where } 
    \beta_{\tau}(t;\pmb{\lambda}) = \sum^J_{j=1}\pi_j(t)\beta^{*(j)}_{\tau}Z^{*(j)}_t(\tau)
\label{case4}
\end{split}
\end{align}
with the transitional probability matrix of $\{Z^{*(j)}_t(\tau), 0 \leq \tau \leq t \}$ to be
 \begin{align*}
 \begin{bmatrix}
    1 & 0  \\
    \rho^{(j)}(\tau,t;\pmb{\lambda}) & 1-\rho^{(j)}(\tau,t;\pmb{\lambda}) \\
\end{bmatrix}
 \text{ and } \rho^{(j)}(\tau,t;\pmb{\lambda}) = 
\frac{1}{1+e^{\lambda^{(j)}_0+\lambda^{(j)}_1\tau}}.
\end{align*} 

Note that this example replaces the indicators $I\big(t \in (t_{j-1}, t_j]\big)$ in model (\ref{case3}) by $\pi_j(t)$. It formulates that a proportion of the response $Y$ at time $t$ resulted from all variants $j$, $j=1,\ldots, J$. Thus, provided that $Y_t = \sum^J_{j=1}Y^{(j)}_t$, one may estimate $\pi_j(t)$ using $Y^{(j)}_t/Y_t$. In general,
the decomposition of $Y_t = \sum^J_{j=1}Y^{(j)}_t$ is unknown.  The unknown $\pi_j(t)$ may be estimated using the possibly available information on the exposure. An analysis of COVID-19 hospitalizations and wastewater viral signals from Ottawa with the model in this example will be presented in Section 3.3.

\section{Analysis of wastewater viral signals and COVID-19 hospitalizations from Ottawa, Canada}
This section reports regression analyses with the proposed model specifications given
in the previous section, the models (\ref{case1})-(\ref{case4}).

\subsection{Analysis with the transition probability $\rho(\tau,t)$ to be constant}

\subsubsection{Comparison with the conventional models}
We start by analyzing the SARS-CoV-2 wastewater measurements and COVID-19 hospitalizations data from Ottawa under model (\ref{case1}).
With the proposed model, $\beta_{\tau}(t) = \beta^*_{\tau}Z^*_{\tau}$ for any given $t$ describes how $Y_t$ is associated with $X_{t-\tau}$, and is comparable to the regression coefficients related to delayed exposure effects in the conventional models, say, $\beta_0,...,\beta_L$ with a fixed $L$. 

All the model parameters are estimated using the marginal likelihood of the data $\{(y_t, x_t); t = 1,...,881\}$, the available observations on the response $Y$ (the hospitalization number) and the exposure $X$ (the COVID-19 viral signals). We assume that $Y_t$ conditional on $\mathcal{H}_x(t), \{z^*_t(\tau), 0 \leq \tau \leq t\}$ are independent for $t = 1,...,881$ and follow the Normal distributions with the means specified in model (\ref{case1}) and variance $\sigma^2$. Let $\pmb{\theta} = (\pmb{\beta^*}, \alpha_0, \sigma, \rho)$ denote the collection of all unknown fixed parameters; $[y_t |\mathcal{H}_x(t), l_t;\pmb{\beta^*}, \alpha_0,\rho, \sigma]$, the contribution of the observation $y_t$
given $\mathcal{H}_x(t)$ and $L(t)= l_t$. Since $l_t$ is not observable, the contribution to the likelihood function of the available observation at time $t$ is
\begin{align*}
L_t(\pmb{\theta}) 
& = \sum_{l=0}^t\big[Y_t|\mathcal{H}_x(t), L(t) = l;\pmb{\beta^*}, \sigma, \alpha_0\big]\big[L(t)=l|\rho \big] \nonumber\\ 
& = \sum_{l=1}^t[Y_t|\mathcal{H}_x(t), L(t)=l;\pmb{\beta^*}, \sigma, \alpha_0]\rho(1-\rho)^{l}.
\end{align*}
The likelihood function of the observed data is $L(\pmb{\theta}) = \prod^{T}_{t = 0}L_t(\pmb{\theta})$.
In practice, one can also consider truncating the lasting time at a sufficiently large upper bound based on the domain knowledge to reduce computational intensity. For instance, we initially truncated the lasting time at 50 days and later reduced it to 30 days, to allow the duration sufficiently captures the critical window of exposure effects for wastewater data analysis.

We obtain the MLE (maximum likelihood estimator) by maximizing the log-likelihood function $log(L_t(\pmb{\theta}))$ with
respect to $\pmb{\theta}$. Specifically, we employ the $nlminb$ function of the \texttt{R} software package to perform the optimization, and set the constraint that $\pmb{\beta}^*$ is non-negative since there is a strong belief that the higher wastewater viral signal would not reflect fewer hospitalizations. And the $hessian$ function in the $numDeriv$ package is used to evaluate the hessian matrix of the log-likelihood function. The covariance matrix of the fixed parameters is then estimated by the negative inverse of the Hessian matrix.

\begin{figure}[hbt!]

\begin{subfigure}{.475\linewidth}
  \includegraphics[width=\linewidth]{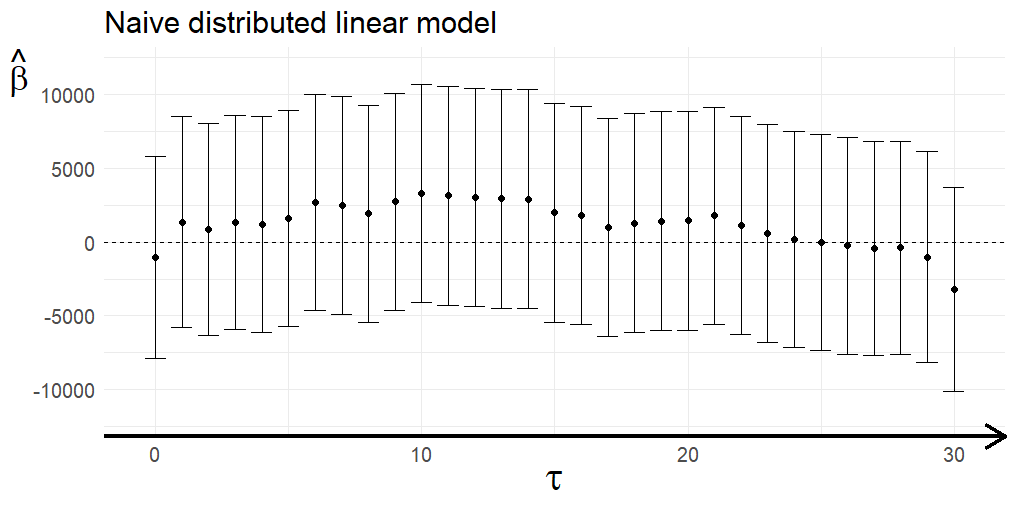}
  \caption{$ E(Y_t|\cdot) = \alpha_0 + \sum_{\tau=0}^{30}\beta_{\tau}X_{t-\tau}$}
  \label{c1}
\end{subfigure}\hfill 
\begin{subfigure}{.475\linewidth}
  \includegraphics[width=\linewidth]{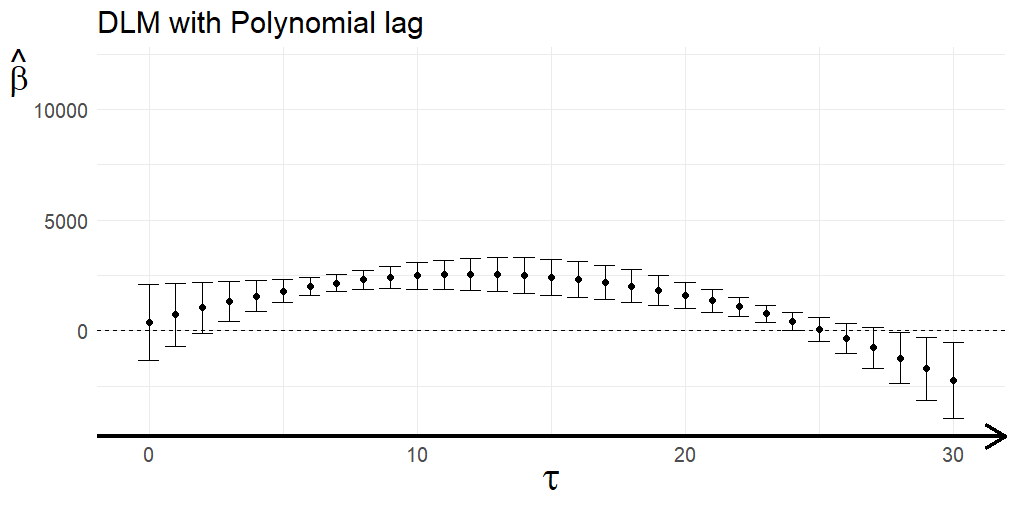}
  \caption{$E(Y_t|\cdot) = \alpha_0 + \sum_{\tau=0}^{30}g_{\beta}(\tau)X_{t-\tau}$}
  \label{c2}
\end{subfigure}

\medskip 
\begin{subfigure}{.475\linewidth}
  \includegraphics[width=\linewidth]{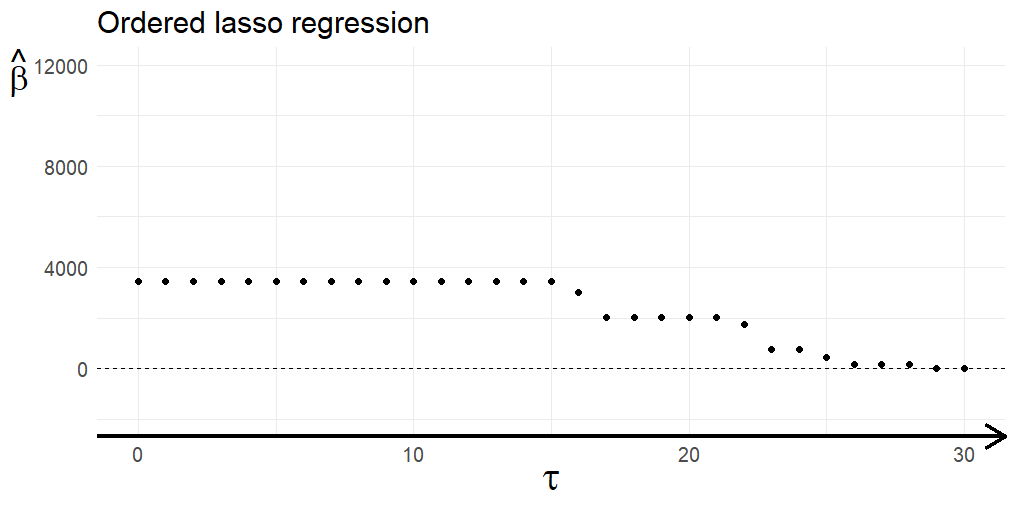}
  \caption{$ E(Y_t|\cdot) = \alpha_0 + \sum_{\tau=0}^{30}\beta_{\tau}X_{t-\tau}$ with constrain $|\beta_0|\geq|\beta_1|\geq...\geq|\beta_l|$}
  \label{c3}
\end{subfigure}\hfill 
\begin{subfigure}{.475\linewidth}
  \includegraphics[width=\linewidth]{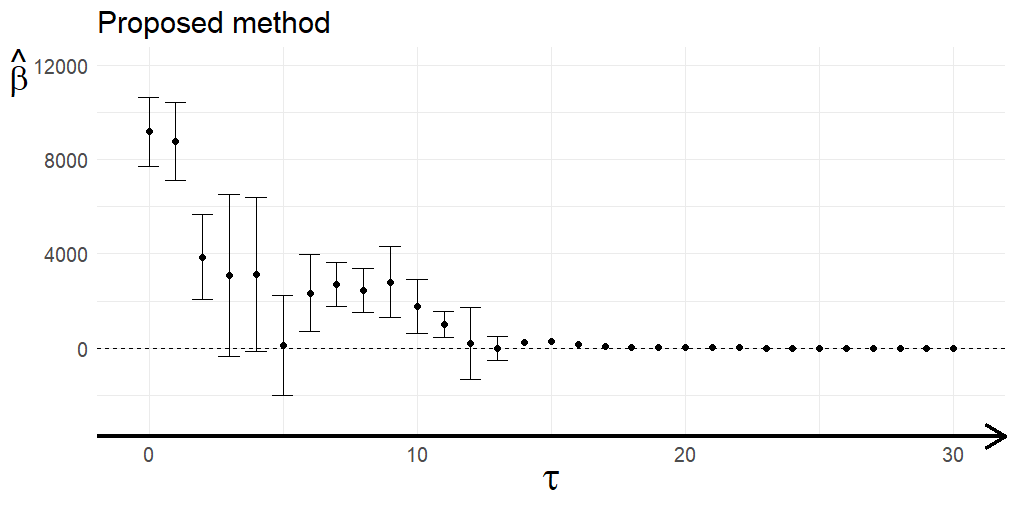}
  \caption{Proposed model (6)}
  \label{p1}
\end{subfigure}

\caption{Comparing proposed model with conventional approaches for exploring the association between COVID-19 hospitalizations and wastewater viral signal.}
\label{fig:roc}
\end{figure}

Figures \ref{fig:roc}a to \ref{fig:roc}c present the estimated $\pmb{\beta}$ obtained from the three conventional methods: the distributed lag linear model, the distributed lag linear model with polynomial lag, and the distributed lag linear model with ordered lasso estimation with a fixed lasting time as $30$ days. Figure \ref{fig:roc}d shows the estimated $\pmb{\beta}$ by the proposed approach: $\widehat{E(\beta_{\tau}|\rho)} = \hat{\beta}^*_{\tau} \hat{E}(Z^*_{\tau}|\hat{\rho}) = \hat{\beta}^*_{\tau} P(L \geq \tau|\hat{\rho})$. An apparent advantage of our proposed model is that it does not require to specify a fixed lasting time, which is needed for all the other methods. By not imposing functional restrictions on $\pmb{\beta}$, the proposed model allows for greater flexibility in detecting the critical window of exposure effects. Our learning from the 
analysis indicates that the critical window is from time lag 0 to 2 days and from time lag 6 to 11 days.

\subsubsection{Empirical Bayes estimation of the lasting time $L(t)$}

\begin{figure}[!htp]
     \centering
		\includegraphics[width=1\linewidth]{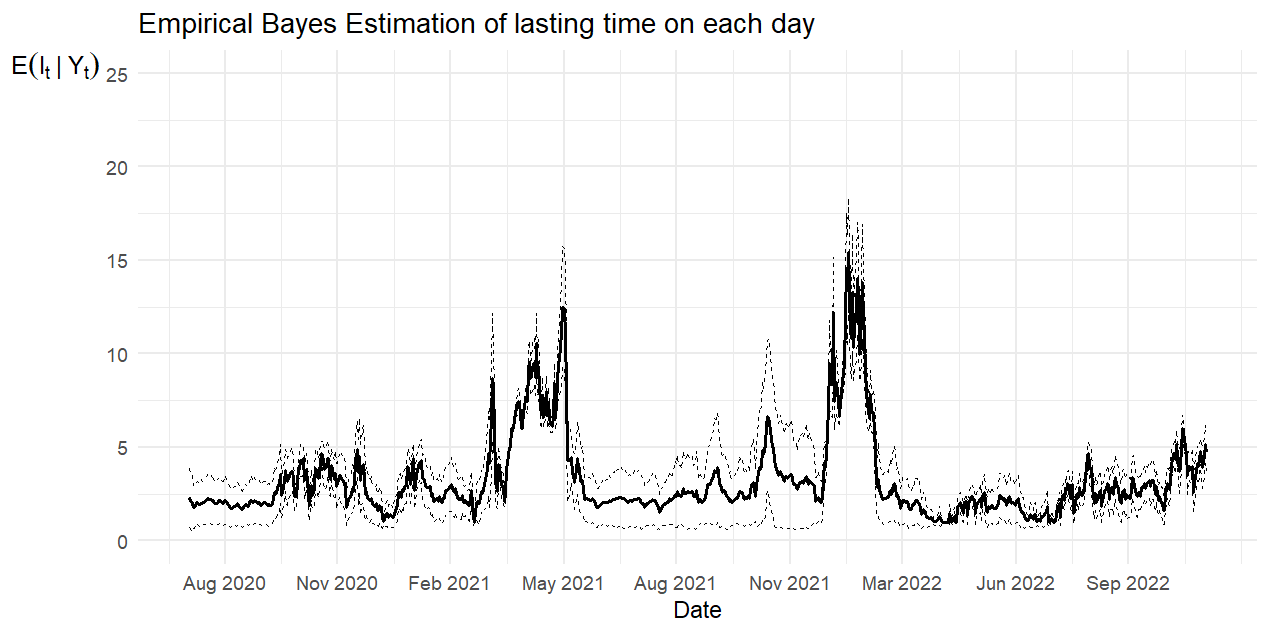}
  \caption{Estimated expectation of the posterior distribution of $(l_t, t=1,...,T)$.}
  \label{dailyLt}
		\end{figure}

Based on our model formulation, the lasting time is random, varying with $t$, and unobservable. Provided with the parameters $(\pmb{\beta^*}, \alpha_0, \sigma, \rho)$, we can obtain the full conditional distribution (posterior) of  the lasting
time associated with time $t$ when the prior distribution of $L(t)$ is $g(\cdot;\rho)$:
\begin{align}\label{postdn}
P(L(t) = l^*|Y_t, \mathcal{H}_x(t)) = \frac{\big[Y_t|\mathcal{H}_x(t), L(t) = l^*;\pmb{\beta^*}, \sigma, \alpha_0\big]
g(l^*;\rho)}{\sum^{t}_{l=0}\big[Y_t|\mathcal{H}_x(t), L(t) = l;\pmb{\beta^*}, \sigma, \alpha_0\big]g( l;\rho)}.
\end{align}
By plugging in $(\pmb{\hat{\beta^*}}, \hat{\alpha}_0, \hat{\sigma}, \hat{\rho})$ into the posterior distribution (\ref{postdn}), we can calculate the expected value $E(L(t)|Y_t)$ as a prediction of $L(t)$ and its variance $Var(L(t)|Y_t)$ on each day $t$. Figure \ref{dailyLt} displays the $E(L(t)|Y_t) \pm Var(L(t)|Y_t)$ over the study period. The changes in $E(L(t)|Y_t)$ from day to day indicate that it is not plausible assume a fixed lasting time for this application. We observed a significant heterogeneity in the lasting time across different pandemic periods. For example, the lasting times during the alpha period (2021 March 23rd to 2021 July 30th) are longer than those during the initial period (before 2021 March 23rd). This learning motivates us to consider the indicators of the different periods as covariates.

\subsection{Analysis with the transition probability $\rho(\tau,t)$ to be time-varying}
According to the preliminary analysis reported in the previous section, the distribution of the lasting time is likely varying across different periods. This guides us to analyze the data with model (\ref{case2}) considering 6 different periods: Initial (before 2021 March 23rd), Alpha (2021 March 23rd to 2021 July 30th), Delta (2021 July 31st to 2021 Dec 20th), Omicron BA.1 (2021 Dec 21st to 2022 March 20th), Omicron BA.2 (2022 March 21st to 2022 June 1st), and Omicron-BA.3+ (after 2022 June 1st). Taking the initial period as the baseline, let $\pmb{w}_t = (w_{1t}, w_{2t},...,w_{5t})$ be the indicators of Alpha, Delta, Omicron BA.1, Omicron BA.2, Omicron BA.3+ period, respectively.

We consider the likelihood based estimation for all parameters based on the marginal likelihood of the observed data $\{(y_t, x_t, \pmb{w}_t); t = 1,...,881\}$. And assume $Y_t$ conditional on $\mathcal{H}_x(t), \pmb{w}_t, \{Z_t^*(\tau), 0 \leq \tau \leq t\}$ for $t = 1,...,881$ are independent and follow the Normal distributions with the means specified in (\ref{case2}) and variance $\sigma^2$. Now the likelihood contribution from one observation $(y_t, \mathcal{H}_x(t), \pmb{w}_t, l_t)$ can be written as $[y_t|\mathcal{H}_x(t), \pmb{w}_t, l_t;\pmb{\beta^*}, \alpha_0,\pmb{\alpha},\pmb{\lambda}, \sigma]$ given $l_t$. Integrating $l_t$ out yields the marginal likelihood function of one observation as $L_t(\pmb{\theta}) 
= \sum_{l=0}^t\big[Y_t|\mathcal{H}_x(t), L(t) = l;\pmb{\beta^*}, \sigma, \alpha_0, \pmb{\alpha}\big]\big[L(t)=l|\pmb{w}_t; \pmb{\lambda}\big]$, which is \\
$\sum_{l=0}^t[Y_t|\mathcal{H}_x(t), L(t)=l;\pmb{\beta^*}, \sigma, \alpha_0, \pmb{\alpha}]\rho(l| \pmb{w}_t; \pmb{\lambda})\prod_{i=0}^{l-1}(1-\rho(i|\pmb{w}_t; \pmb{\lambda}))$. Adopting the convention that $\prod_{i=0}^{-1}(1-\rho(i|\pmb{w}_t; \pmb{\lambda})) = 1$, the likelihood function of the observed data is then $L(\pmb{\theta}) = \prod^{T}_{t = 0}L_t(\pmb{\theta})$.

\begin{figure}[!htp]
\centering
  \includegraphics[width=0.7\linewidth]{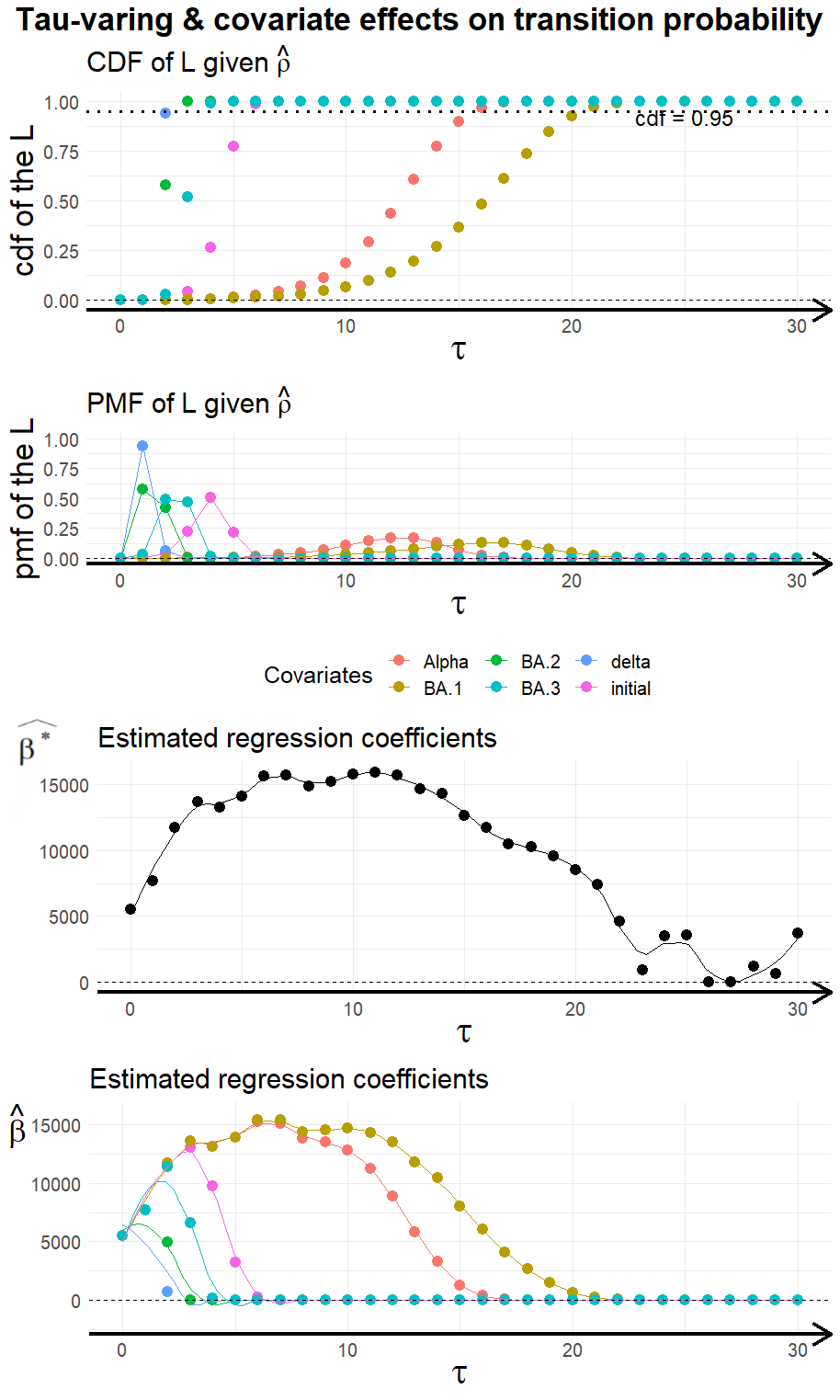}
\caption{Parameter estimates of probability mass function of $L = max\{\tau; Z^*_{\tau} = 1\}$; cumulative distribution function of $L$; fixed regression coefficients $\hat{\pmb{\beta}^*}$; estimated expectation of $\hat{E}(\beta(\tau|\pmb{W};\pmb{\lambda})) = \hat{\beta^*_{\tau}} E(Z^*_{\tau}|\pmb{W};\pmb{\hat{\lambda}})$ for full model considered in Section 4.2.}
\label{fig:test}
\end{figure}

Figure \ref{fig:test} contains four sub-figures that present the following: the estimated cumulative distribution function (CDF) of the lasting time $L(t)$, the estimated probability mass function (PMF) of $L(t)$, the estimated $\hat{\beta}^*_{\tau}$, and $\hat{E}(\beta(\tau|\pmb{W};\hat{\pmb{\lambda}}))$ across time lag $\tau$, respectively. Different covariates (for time periods) are color-coded in the 1st, 2nd, and 4th sub-figures, while the 3rd sub-figure shows a shared $\pmb{\beta}^*$ for all periods. Smoothed curves by locally estimated scatterplot smoothing (loess) in the 2nd, 3rd, and 4th sub-figures are added for the visualization purposes. 

Table \ref{tableexample2} presents the parameter estimates for $(\alpha_0, \pmb{\alpha},\pmb{\lambda})$ for the full model. The results suggest that the order of the lasting times of the delayed association between COVID-19 hospitalizations and wastewater viral signals, from longest to shortest, be: BA.1 period, Alpha period, initial period, BA.3 period, BA.2 period, and Delta period. Significant differences are detected between the following periods: 
\begin{itemize}
    \item BA.1 and Alpha period (p-value $<$ 0.05, $H_0: \lambda_2 = \lambda_4$).
    \item Alpha and initial period (p-value $<$ 0.05, $H_0: \lambda_1 = \lambda_2$)
    \item Initial versus BA.3 period (p-value $<$ 0.05, $H_0: \lambda_1 = \lambda_6$)
    \item BA.3 versus BA.2 period (p-value $<$ 0.05, $H_0: \lambda_5 = \lambda_6$). 
\end{itemize}
Note that significant differences were tested only for sequentially adjacent periods depending on the order of their lasting times.

\subsection{Two stratification strategies}
The analysis with pooled data may not fully capture the varying associations between COVID-19 hospitalizations and wastewater viral signals under the model that assumes they differ only in the distribution of lasting times but share the same underlying $\pmb{\beta}^*$ across different periods.  For instance, during the Omicron BA.1 period, the effect size appears large since a low level of wastewater viral signals corresponds to a large number of hospitalizations. However, under the assumption that the underlying $\pmb{\beta}^*$ is the same for all periods, the model can only account for the high number of hospitalizations by suggesting a longer-lasting time, allowing more wastewater viral signals to contribute to predicting hospitalizations for the Omicron BA.1 period. To address the limitation, we propose two stratification strategies and examplify them in the following.

The arrival of the Omicron variant in November 2021 brought with a more transmissible virus at the time when population immunity was boosted by recent vaccination campaigns in Ottawa. We first consider stratified analysis for pre-Omicron and Omicron periods, splitting on 2021 Nov 26th as adopted in practice. This strategy is referred to as ``\textit{Hard}" stratification using model (\ref{case3}).

However, there is always a mixture of multiple variants during the switching period between two VOCs. The mixture indicates a lack of a definitive cut-off point between the two periods. To account for it, we consider a ``\textit{Soft}" stratification approach, using model (\ref{case4}) to compare and capture the process of the new variant's dominance. In Section 2 of the supplementary file, we motivate this strategy further by displaying the proportions of different variants in Ottawa's wastewater.

To improve the efficiency of the parameter estimating, we further assume all COVID-19 hospitalizations are related to the pre-Omicron variant before 2021 Nov 1st; after 2022 Jan 31st, all hospitalizations are related to the Omicron variant. Between 2021 Nov 1st and 2022 Jan 31st, the hospitalizations represent a mixture of pre-Omicron and Omicron caused cases. The recorded proportions of pre-Omicron and Omicron wastewater viral signals are displayed in Section 2 of the supplementary. The corresponding viral signals, $\pmb{X}^{(1)}$ for pre-Omicron and $\pmb{X}^{(2)}$ for Omicron, can be calculated for the two periods, while only the aggregated hospitalization counts $\pmb{Y} = \pmb{Y}^{(1)}+\pmb{Y}^{(2)}$ is avaliable.

\begin{figure}[!htp]
\centering
\includegraphics[width=.48\textwidth]{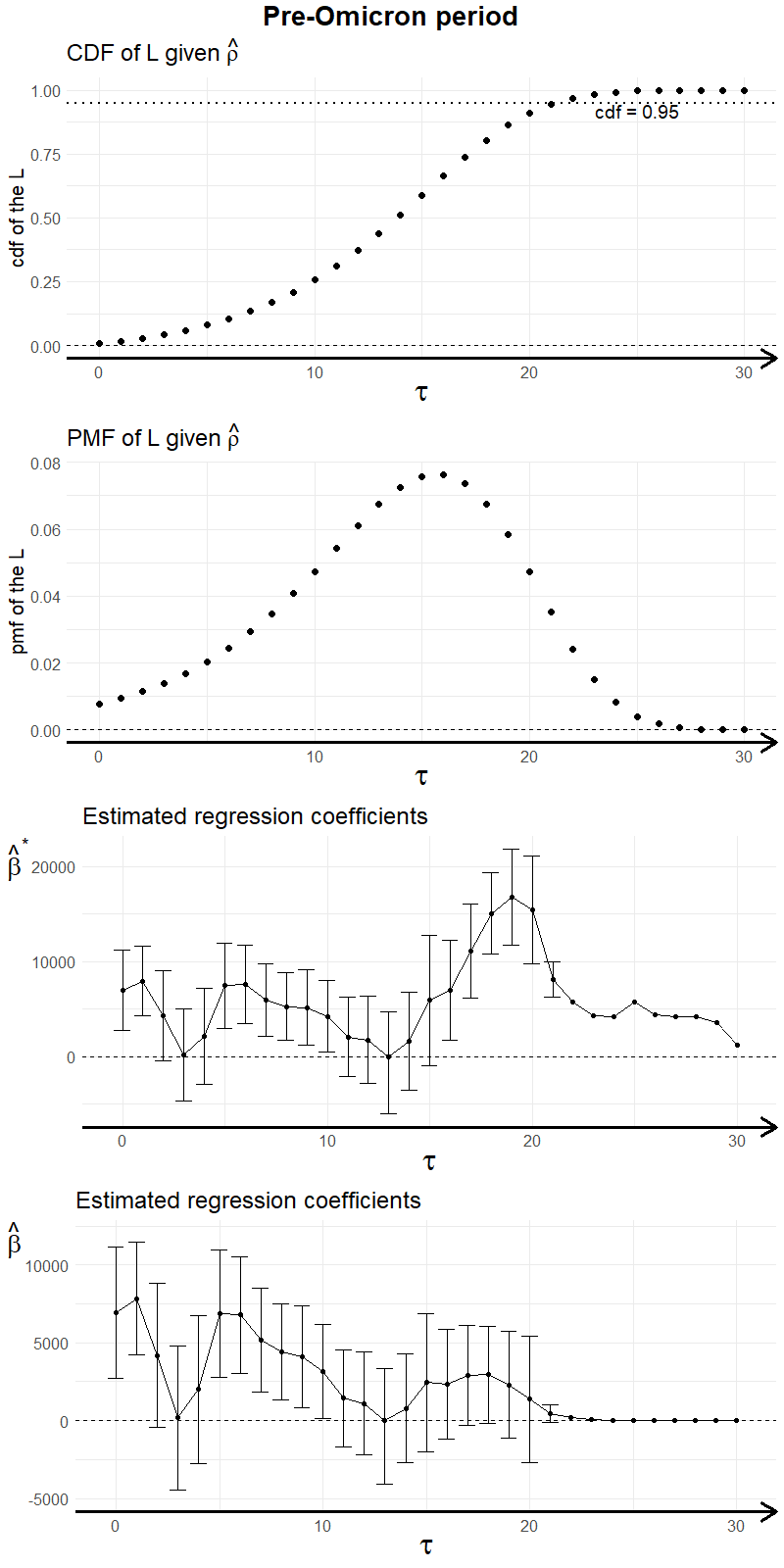}
\includegraphics[width=.48\textwidth]{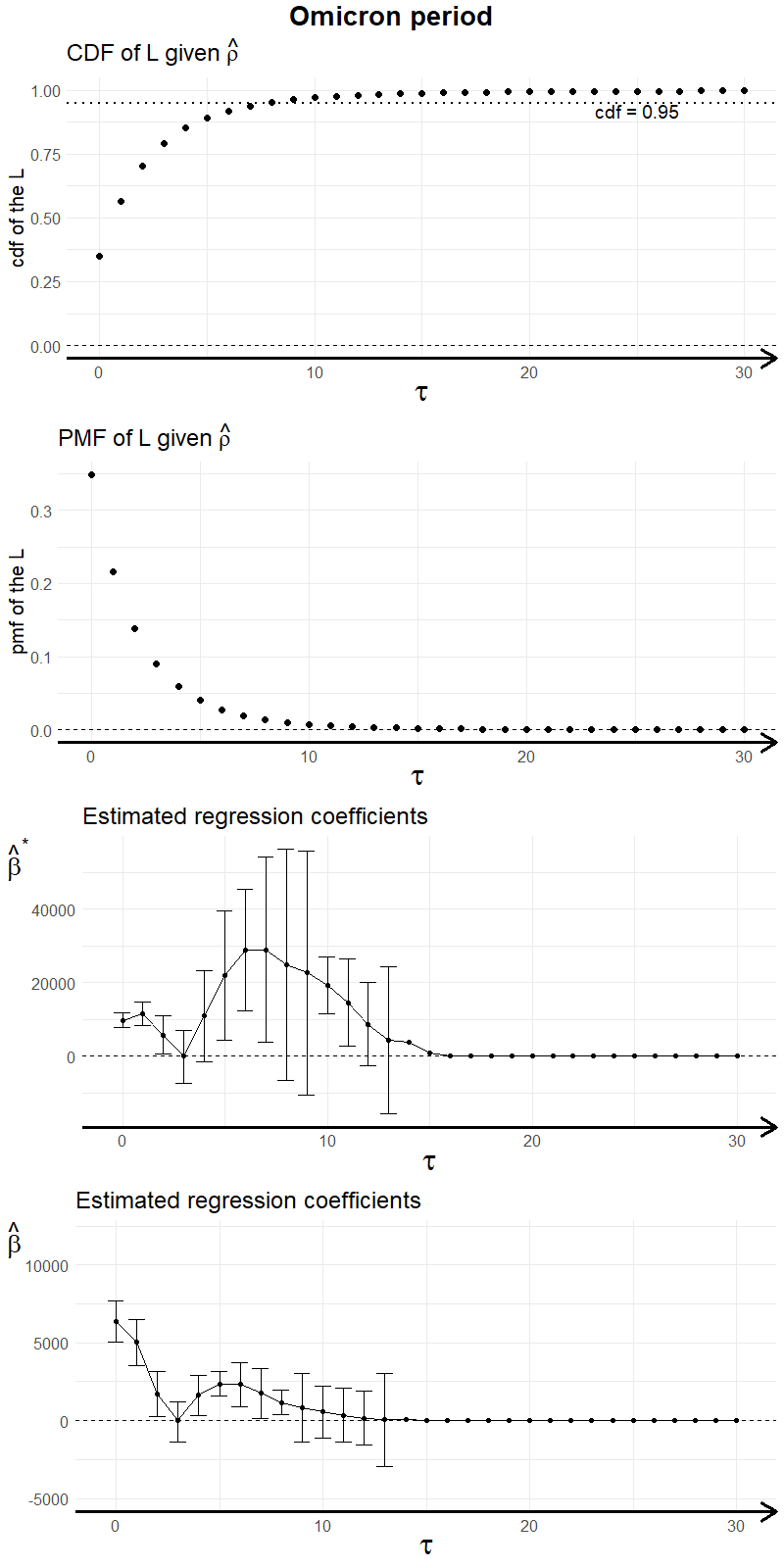}
\caption{First row shows estimated cumulative distribution functions (CDF) of lasting time for pre-Omicron and Omicron period; second row shows estimated probability mass functions (CDF); third row shows the estimated underlying $\pmb{\beta}^*$; forth row shows the estimated $E(\pmb{\beta})$.}
\label{stratified1}
\end{figure}

Figure \ref{stratified1} presents the results of estimated CDF, PMF, $\pmb{\beta}^*$, and $E(\pmb{\beta})$ for both pre-Omicron and Omicron periods, and the parameter estimates of $(\lambda_0^{(1)},\lambda_1^{(1)},\lambda_0^{(2)},\lambda_1^{(2)})$ are tabulated in Table \ref{tablestratified1}. Significant differences are found between the pre-Omicron and Omicron periods in terms of both the distribution of lasting time $L(t)$ and the underlying $\pmb{\beta}^*(t)$. Specifically, the results indicate that the pre-Omicron has a longer lasting time on average, with critical windows of exposure effects identified as day 0 to 1 and day 5 to 10. The critical windows of the Omicron period are days 0 to 2 and days 4 to 8. 

Notably, model (\ref{case3}) and (\ref{case4}) yield similar results, as the estimated $\pi(t)$ in ``\textit{Soft}" stratified analysis quickly approaches 1. This finding supports the understanding that the transition between the pre-Omicron and Omicron periods is rather rapid, and therefore, the ``\textit{Hard}" stratification works well for these two strata. We conducted an additional comparison using "\textit{Hard}" and "\textit{Soft}" stratified analyses, focusing on the transition from the Omicron BA.1 period to the Omicron BA.2+ period. This comparison reveals additional findings from "\textit{Soft}" stratification, with a detailed explanation provided in Section 2.2 of the supplementary file.

\section{Discussion}
In this paper, we propose a class of Markov-modulated models with a random lasting time to study how the response is associated with an exposure. Under the model formulation, the lasting time of the delayed effect is modeled as a random variable governed by a hidden Markov or semi-Markov chain. This approach addresses a key limitation in traditional distributed lag models, which require the specification of a fixed lasting time, this may be challenging in many applications. We also proposed a ``Soft" stratified analysis when the cut-off between different strata is unknown. With the developed model, we studied the delayed association of COVID-19 hospitalization and wastewater viral signals in Ottawa, Canada. Our proposed learning strategy may lead to reasonable inference about not only how COVID-19 hospitalization is associated with wastewater viral signal, but also how long the association lasts. It is worth noting that the proposed model is a general extension of DLMs and can be adapted in many different fields with the suitable specifications of model components. For instance, as illustrated in Example 1 in Section 2.3, the regression coefficients $\pmb{\beta}^*$ dynamically reduce to 0 without any specifications of functional restrictions on them, this could be useful when the researcher is not confident to specify the functional form of the delayed effects. In addition, our idea has the potential to motivate numerous future investigations in the field of distributed lag models. For example, in the continuous-time case, one might consider extending historical functional regression with a continuous-time Markov process. In the spatio-temporal context, exploring random ``persistence distance" could also be valuable.

The parameter estimation of the proposed model is carried out by maximizing the marginal likelihood of the observed data. Under the model assumptions, integrating the whole hidden random process is equivalent to integrating the lasting time $L$, and it is tractable. However, we note that the marginal likelihood function of one observation is the sum of $t$ terms which can cause the optimization to be computationally intensive. In practice, one can limits the maximal time lag to a sufficient large number and has reduce the length of regression coefficients $\pmb{\beta}^*$. When a longer maximal time lag is needed, a more efficient estimation procedure may need to be investigated. The model we propose can be more flexible if considering the response variance varying overtime. Further research includes to investigate more types of variants and study the general changes when new variant arrives, which could be useful for predicting hospitalizations in the future.

\section*{Acknowledgement}
The study was supported by the CANSSI (Canadian Statistical Sciences Institute) Collaborative Research Team (CRT) in ``The Application of Statistical Methods to Wastewater Analysis". The authors thank Dr. Tyson Graber, a researcher at the Children's Hospital of Eastern Ontario Research Institute, for providing the data and valuable discussions on the proportions of different variants observed in Ottawa's wastewater measurements.

\clearpage

\appendix

\section{Figures}

\clearpage

\clearpage
\section{Tables}

\begin{table}[!htb]
\centering
\begin{threeparttable}
\centering
            \begin{tabular}{l r r}
			    \toprule
                \textbf{Parameter} & \textbf{Estimates ($SE$)}   \\
			\midrule
	 	  Intercept (initial period) ($\alpha_0$)   &  5.671 (1.626) \\
               \tnote{a}  Indicator of Alpha period ($\alpha_1$)  &  \pmb{-17.178 (3.157)} \\
               \tnote{a} Indicator of Delta period ($\alpha_2$)  &  6.115 (1.777) \\
               \tnote{a} Indicator of Omicron-BA.1 period ($\alpha_3$)  &  \pmb{-11.344 (2.475)} \\
               Indicator of Omicron-BA.2 period ($\alpha_4$)  &  5.157 (2.721) \\
               \tnote{a} Indicator of Omicron-BA.3+ period ($\alpha_5$) &  7.962 (2.095) \\
              \midrule
              Intercept of $\rho(\tau;\pmb{\lambda})$  ($\lambda_0$) & \pmb{7.244 (0.149)} \\
              \tnote{b} Change rate during initial ($\lambda_1$) & \pmb{-2.012 (0.227)}\\
              \tnote{b} Change rate during Alpha ($\lambda_2$) & \pmb{-0.534 (0.020)}\\
              \tnote{b} Change rate during Delta ($\lambda_3$)  & \pmb{-9.978 (2.499)}\\
              \tnote{b} Change rate during Omicron-BA.1 ($\lambda_4$)  & \pmb{-0.383 (0.031)}\\
              \tnote{b} Change rate during Omicron-BA.2 ($\lambda_5$)  & \pmb{-7.563 (0.804)}\\
              \tnote{b} Change rate during Omicron-BA.3+ ($\lambda_6$) & \pmb{-3.628 (0.510)}\\
              \bottomrule
		\end{tabular}
\begin{tablenotes}\footnotesize
\item [a] Bold estimates of regression coefficients $\alpha_1$ to $\alpha_5$ indicate the estimate is significantly different from $\alpha_0$, which implies the corresponding period has significant different intercept compare to initial period.
\item [b] Bold estimates of $\lambda_1$ to $\lambda_6$ indicate the estimate is significantly different from 0, which implies the semi-Markov process is needed rather than Markov process for model $L$'s distribution. Because $\lambda_s = 0$ indicates $\{Z^*_{\tau}, \tau \geq 0\}$ is a Markov chain during $sth$ period. 
\end{tablenotes}

                \caption{Estimates of the parameters $(\alpha_0,\pmb{\alpha}, \pmb{\lambda})$ for full model considered in Section 4.2.}
\label{tableexample2}
\end{threeparttable}
\end{table}


\begin{table}[!htb]
\centering
\begin{threeparttable}
\centering
            \begin{tabular}{l r r}
			   \toprule
			   & \textbf{Pre-Omicron period} & \textbf{Omicron period}  \\

                  \textbf{Parameter} & \textbf{Estimates ($SE$)}& \textbf{Estimates ($SE$)}   \\
			   \midrule
	 	  Intercept ($\alpha$) & 2.977 (.637) &  14.069 (.763) \\
               Standard deviation  ($\sigma$) & 6.121 (.336) &   7.935 (.421) \\
            \midrule
             Intercept of $\rho(\tau;\pmb{\lambda})$  ($\lambda_0$) & 4.867 (.327) &  0.627 (.085)\\
             \tnote{a} Change rate of $\rho(\tau;\pmb{\lambda})$ ($\lambda_1$) & \pmb{-0.211 (.021)} & \pmb{0.072 (.020)} \\
              \bottomrule
		\end{tabular}
\begin{tablenotes}\footnotesize
\item [a] Bold estimates of $\lambda_1$ indicate significance deviating from 0, which indicate the importance of the $\tau$-varying transition probability.
\end{tablenotes}

                \caption{Estimates of the parameters $(\alpha, \sigma, \pmb{\lambda})$ for stratified analysis in Section 3.3. The pre-omicron period in Ottawa is from 2020 June 16th to 2021 Nov 26th; the Omicron period in Ottawa is from 2021 Nov 27th to 2022 Nov 13rd. The stratification is a one-to-one correspondence with the stratification in Figure \ref{stratified1}.}
\label{tablestratified1}
\end{threeparttable}
\end{table}
\clearpage

\clearpage
\bibliographystyle{agsm}
\bibliography{Bibliography-MM-MC}

\end{document}


\def\spacingset#1{\renewcommand{\baselinestretch}%
{#1}\small\normalsize} \spacingset{1}


\if1\blind
{
  \title{\bf Supplementary Materials: Learning associations of COVID-19 hospitalizations with wastewater viral signals by Markov modulated models}
  \author{K. Ken Peng,
    Charmaine B. Dean, 
    Robert Delatolla and 
    X. Joan Hu \\
   }
  \maketitle
} \fi

\if0\blind
{
  \bigskip
  \bigskip
  \bigskip
  \begin{center}
    {\LARGE\bf Exploring associations of COVID-19 hospitalizations with wastewater viral signals by Markov modulated model with distributed lag}
\end{center}
  \medskip
} \fi

\bigskip

\spacingset{1.9} 

\section{Supplementary material for Section 2.3 in main text - Example S1: ${Z^*_t(\tau), 0 \leq \tau \leq t }$ as a Markov chain conditional on different periods}

Considering $\beta^*_{\tau}(t)$ as fixed unknown parameters $(\beta^*_0,...,\beta^*_t)$. For the ``stochastic" part, consider $\rho(\tau, t) = \rho(t|\pmb{\lambda})$ as a function of $t$, where $\pmb{\lambda}$ is the related parameter vector. We formulate $\rho(t|\pmb{\lambda})$ as a step function of time $t$, further specifying it as:
\begin{equation*}
 \rho((\tau, t);\pmb{\lambda}) = \begin{cases}
 \rho_0(\tau|\pmb{\lambda}) = \frac{1}{1+e^{\lambda_0}}, & t \in (t_0,t_1)\\
 \rho_1(\tau|\pmb{\lambda}) = \frac{1}{1+e^{\lambda_0+\lambda_1}}, & t \in (t_1,t_2)\\
... 
\end{cases}
\end{equation*}
where $\{t_j, j = 1,...,J\}$ split the entire study period into $J$ sub-periods. Denote $t_0$ and $t_J$ as the start and end dates of the study period.

Based on this formulation of $\rho(t|\pmb{\lambda})$, $\{Z^*_t(\tau), 0 \leq \tau \leq t \}$ forms a 2-state Markov chain, and our proposed model can be written as:
\begin{align}
\begin{split}
    & E(Y_t|\mathcal{H}_x(t), \{Z^*_t(\tau), 0 \leq \tau \leq t\}; \alpha_0, \pmb{\alpha}, \pmb{\beta}^*) 
    = \sum^J_{j=1}I\big(t\in(t_{j-1},t_j)\big)\alpha^{(j)}_0 + \sum_{\tau=0}^t\beta( t|\pmb{\lambda})X_{t-\tau} \\
    & \text{ with } 
    \beta(t|\pmb{\lambda}) = \begin{cases}
 \beta^*_{\tau}, & Z^*_t(\tau) = 1\\
0, & Z^*_t(\tau) = 0 \\
\end{cases}   
\label{case2}
\end{split}
\end{align}
with the transitional probability matrix of $\{Z^*_t(\tau), 0 \leq \tau \leq t \}$ is 
\begin{equation*}
P(\tau, t|\pmb{\lambda}) = \begin{bmatrix}
    1 & 0  \\
    \rho(t|\pmb{\lambda}) & 1-\rho(t|\pmb{\lambda}) \\
\end{bmatrix}
\end{equation*}

This model specification serves as an intermediate example, bridging Example 1 and Example 2 in the main text. It considers the distribution of $L$ as being dependent on different periods, while maintaining the assumption that the probability mass function of $L(t)$ monotonically decreases across $\tau$. Figure \ref{example1.5} displays the estimated cumulative distribution function (CDF) of $L$, the estimated probability mass function (PMF) of $L$, $\hat{\beta}^*_{\tau}$, and $\hat{E}(\beta(\tau|\pmb{W};\hat{\pmb{\lambda}}))$ across time lag $\tau$, respectively. 

Table \ref{tableexample1.5} presents the parameter estimates for $(\alpha_0, \pmb{\alpha},\pmb{\rho})$. The results suggest that the order of the lasting times of the delayed association between COVID-19 hospitalizations and wastewater viral signals, from longest to shortest, is: Alpha period, BA.1 period, initial period, BA.3 period, BA.2 period, and Delta period. Significant differences are detected between the following periods:
\begin{itemize}
    \item BA.1 and initial period (p-value $<$ 0.05, $H_0: \rho_3 = \rho_0$)
    \item BA.3 versus BA.2 period (p-value $<$ 0.05, $H_0: \rho_5 = \rho_4$). 
\end{itemize}
Note that significant differences were tested only for sequentially adjacent periods depending on the order of their lasting times.

These results are similar with Section 3.2 of the main text. However, it is notable that this model was unable to detect a significant difference between the Omicron BA.1 period versus the Alpha period, and the initial period versus the BA.3 period, whereas Example 2 in the main text does. Additionally, it is worth noting that the model in Example 2 has a better fit, as indicated by a lower Akaike Information Criterion (AIC), improving from 6800.325 in the Example in the supplementary file to 6465.667, with the inclusion of only one additional parameter.

\section{Supplementary material for Section 3.3 in main text - Additional materials}

\subsection{Proportion of different variants of concern}

Figure \ref{variantsproportion} illustrates the proportion of different variants of concern (VOC) of COVID-19 detected in wastewater over time. Omicron variants are identified starting from ``N: P13L", and the black dashed line denotes the date, November 26, 2021, where we split the pre-Omicron and Omicron periods.

\subsection{Omicron BA.1 versus Omicron BA.2+ period}

Adapting same idea as in Section 3.3 in main text, we compare the ``hard" and ``soft" stratified analyses separating the Omicron BA.1 and Omicron BA.2+ periods. For the ``hard" stratification, we consider Omicron BA.1 period is from 2021 Nov 20th to 2022 March 21st, and Omicron BA.2+ period is from 2022 March 21st to 2022 Nov 13rd. We consider the switch period from BA.1 to BA.2 is from 2022 February 1st to 2022 May 1st. Figure \ref{hardstratified2} presents the results of estimated CDF, PMF, $\pmb{\beta}^*$, and $E(\pmb{\beta})$ for Omicron BA.1 and BA.2+ periods employing model in Example 3 in the main text, Figure \ref{softstratified2} shows results employing model in Example 4 in the main text. Table \ref{tablestratified2} displays the corresponding parameter estimates of $(\lambda_0^{(1)},\lambda_1^{(1)},\lambda_0^{(2)},\lambda_1^{(2)})$. Unlike the stratified analysis for Pre-Omicron and Omicron periods, the comparison between Omicron BA.1 and Omicron BA.2 do presents different results using ``hard" and ``soft" stratified analysis. 

Comparing Figure \ref{hardstratified2} and \ref{softstratified2}, the model suggests BA.1 period has a shorter lasting time, but with larger effect size comparing to BA.2 period. This indicates the COVID-19 hospitalizations response fast after the changes in wastewater viral signals, and a same level of wastewater viral signals could reflect more hospitalizations. The standard error estimates were large when considering ``hard" stratification, this may caused by the short duration of BA.1 period. As for BA.2 periods, both analyses suggests a long lasting time, and uncertain association between COVID-19 hospitalizations and wastewater viral signals which is essentially similar as conventional distributed lag linear model. Notably, the estimated $\pi(t)$, the proportion of Omicron BA.2+ caused hospitalizations does show a more clear trend, which is shown at Figure \ref{hospproportion} in supplementary file.

\clearpage

\appendix

\section{Figures}
\renewcommand{\thefigure}{S\arabic{figure}}

 \begin{figure}[!htp]
\centering
\includegraphics[width=0.65\textwidth]{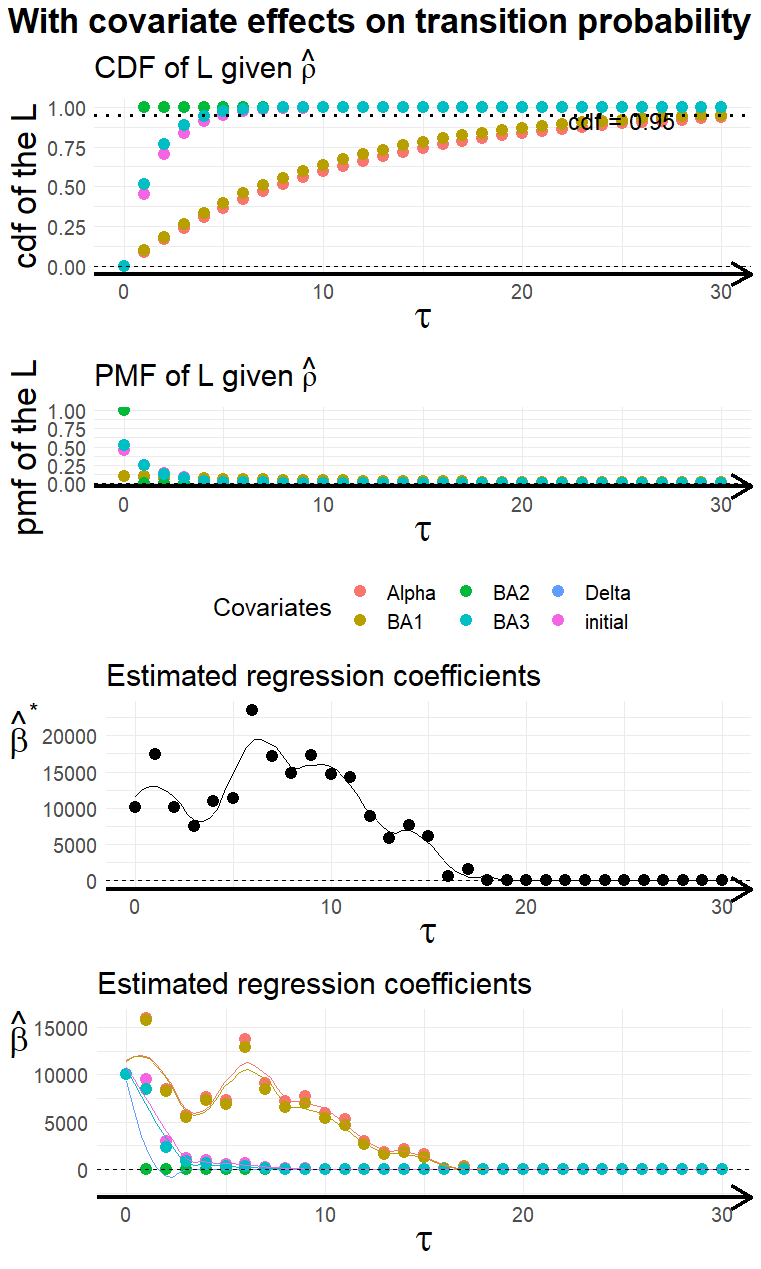}
\caption{Parameter estimates of probability mass function of $L = max\{\tau; Z^*_{\tau} = 1\}$; cumulative distribution function of $L$; fixed regression coefficients $\hat{\pmb{\beta}^*}$; estimated expectation of $\hat{E}(\beta(\tau|\pmb{W};\pmb{\lambda})) = \hat{\beta^*_{\tau}} E(Z^*_{\tau}|\pmb{W};\pmb{\hat{\lambda}})$ for model considered in Example S1.}
\label{example1.5}
\end{figure}

\clearpage

\begin{figure}[!htp]
    \centering
    \includegraphics[width=1\textwidth]{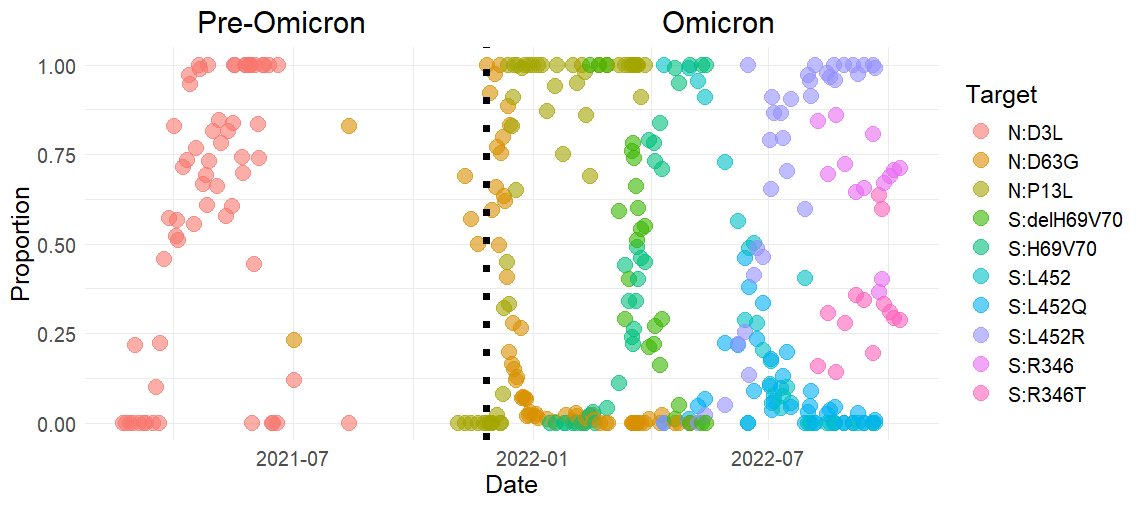}
    \caption{The proportion of different COVID-19 variants within wastewater. Black dashed line is 2021 Nov 26th the date splitting pre-Omicron and Omicron period in hard stratified analysis in the main text.}
    \label{variantsproportion}
\end{figure}

\clearpage

 \begin{figure}[!htp]
\centering
\includegraphics[width=.48\textwidth]{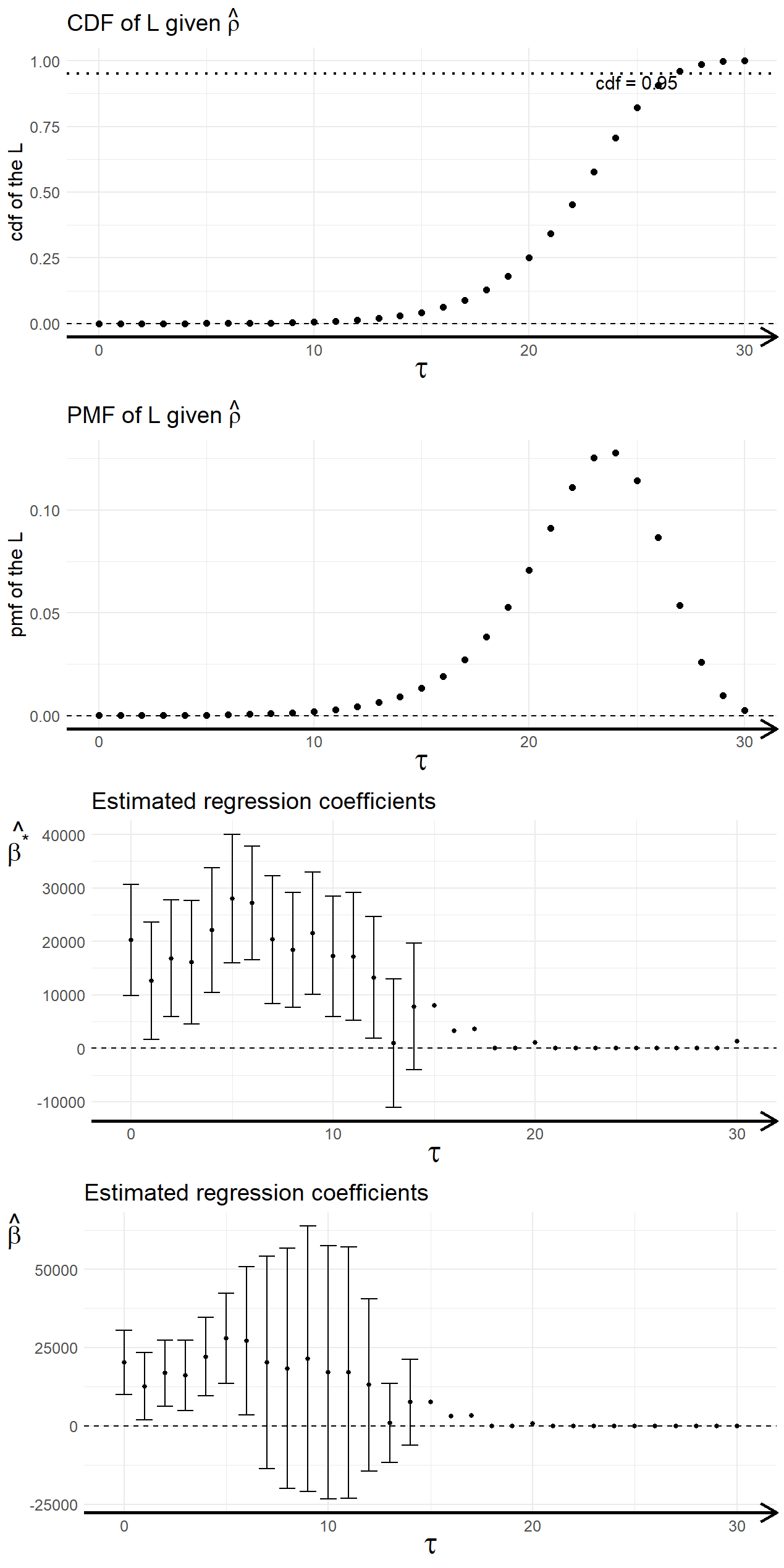}
\includegraphics[width=.48\textwidth]{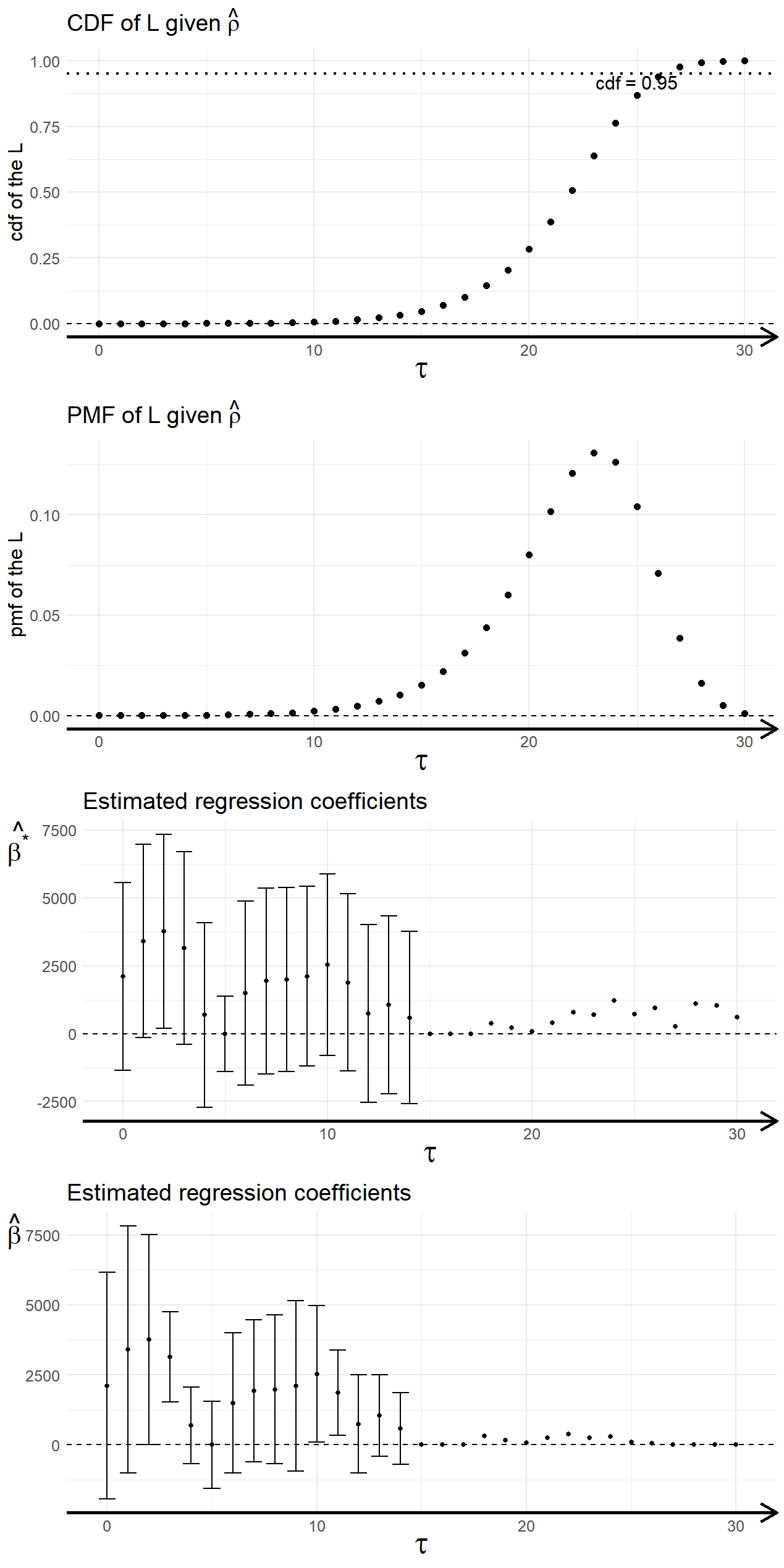}
\caption{``Hard" stratification analysis comparing Omicron BA.1 and Omicron BA.2+ periods. First row shows estimated cumulative distribution functions (CDF) of lasting time for pre-Omicron and Omicron period; second row shows estimated probability mass functions (CDF); third row shows the estimated underlying $\pmb{\beta}^*$; forth row shows the estimated $E(\pmb{\beta})$. The left column corresponds to the BA.1 period, while the right column corresponds to the BA.2+ period.}
\label{hardstratified2}
\end{figure}

\clearpage

 \begin{figure}[!htp]
\centering
\includegraphics[width=.48\textwidth]{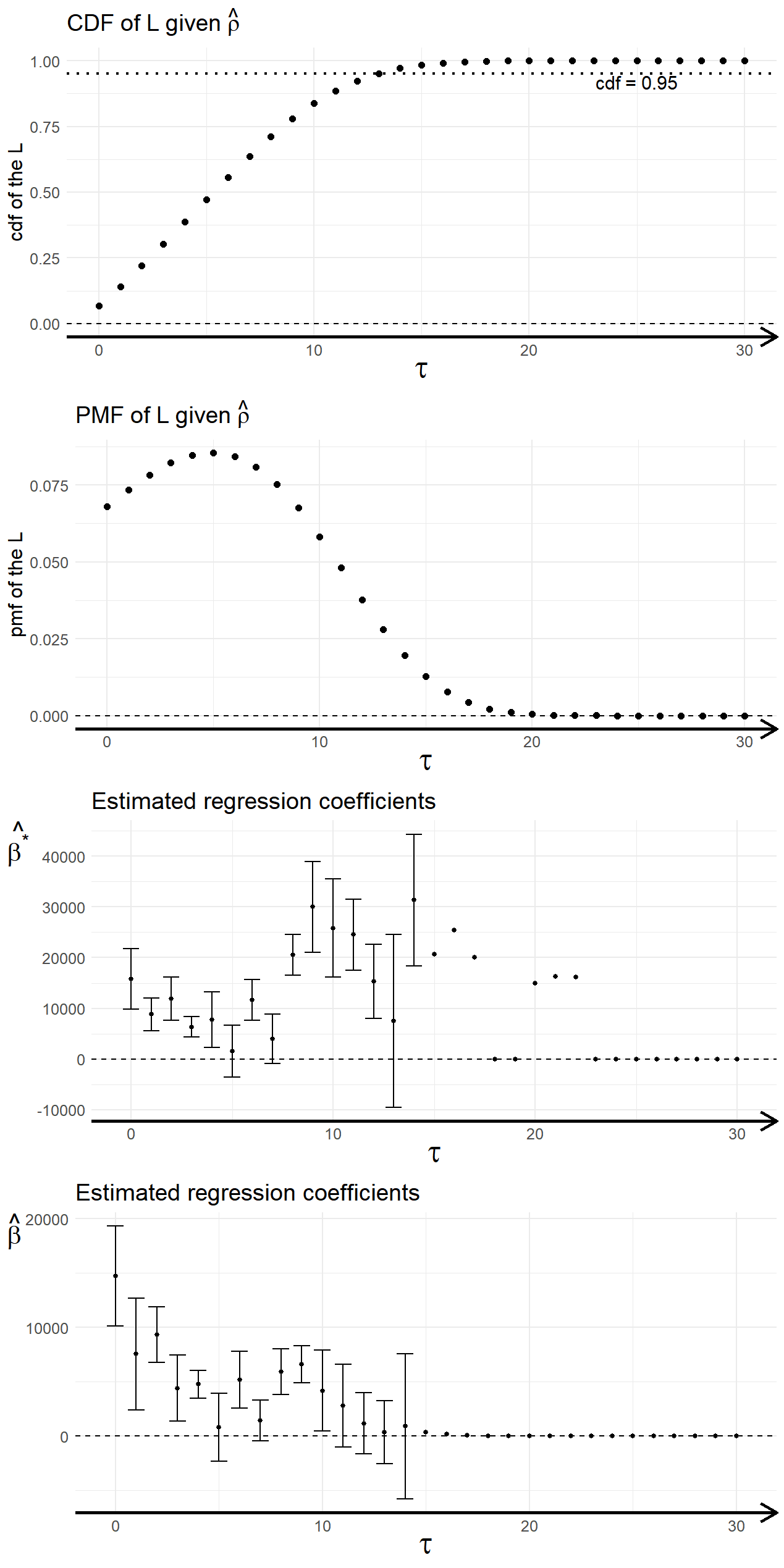}
\includegraphics[width=.48\textwidth]{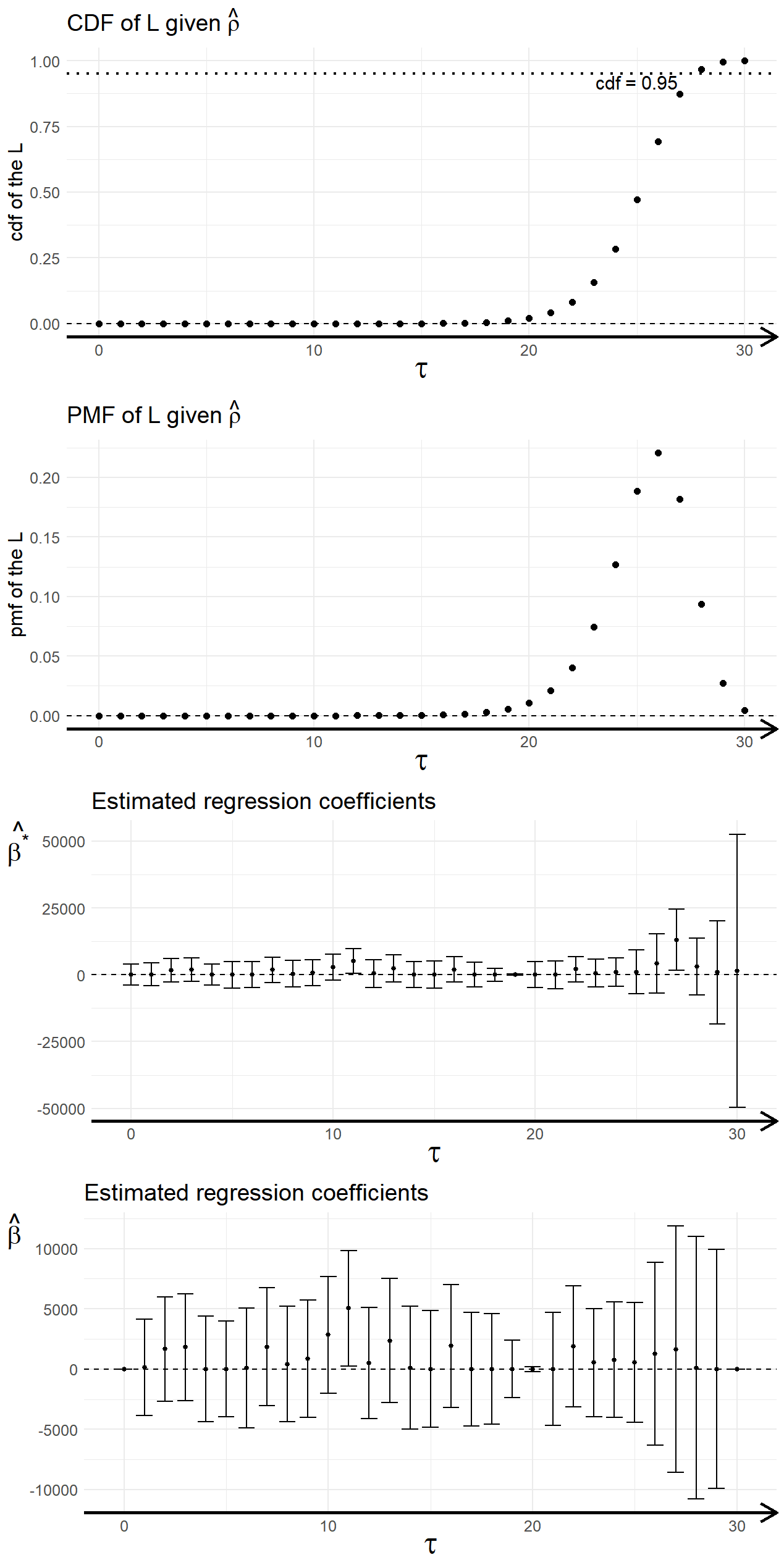}
\caption{``Soft" stratification analysis comparing Omicron BA.1 and Omicron BA.2+ periods. First row shows estimated cumulative distribution functions (CDF) of lasting time for pre-Omicron and Omicron period; second row shows estimated probability mass functions (CDF); third row shows the estimated underlying $\pmb{\beta}^*$; forth row shows the estimated $E(\pmb{\beta})$. The left column corresponds to the BA.1 period, while the right column corresponds to the BA.2+ period.}
\label{softstratified2}
\end{figure}

\clearpage

 \begin{figure}[!htp]
\centering
\includegraphics[width=.8\textwidth]{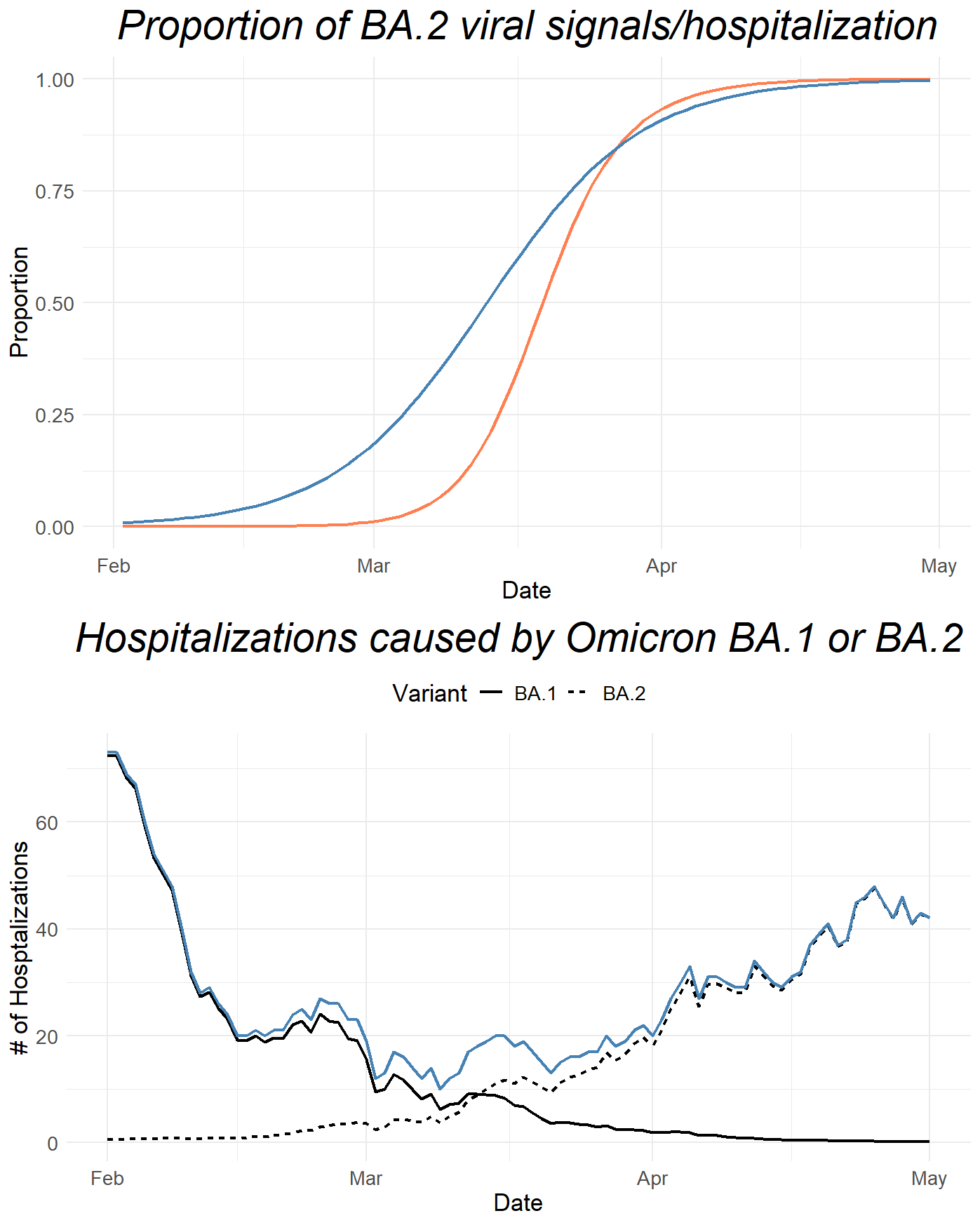}
\caption{Top figure: Estimated proportion of hospitalizations caused by Omicron BA.2+ (blue curve) compared to the observed proportion of Omicron BA.2+ viral signals (orange curve). Bottom figure: Estimated number of hospitalizations attributed to Omicron BA.1 (solid curve) and Omicron BA.2+ (dashed curve), where the blue curve shows the observed aggregated hospitalizations.}
\label{hospproportion}
\end{figure}

\clearpage

\section{Tables}
\renewcommand{\thetable}{S\arabic{table}}

\begin{table}[!htb]
\centering
\begin{threeparttable}
\centering
            \begin{tabular}{l r r}
			    \toprule
                \textbf{Parameter} & \textbf{Estimates ($SE$)}   \\
			\midrule
	 	  Intercept (initial period) ($\alpha_0$) &   \pmb{13.872 (0.869)}   \\
               \tnote{a} Indicator of Alpha ($\alpha_1$) & \pmb{-18.528 (1.560)}    \\
                Indicator of Delta ($\alpha_2$) &  -1.732 (1.085) \\
                Indicator of Omicron-BA.1 ($\alpha_3$) &  1.269 (1.650)  \\
               \tnote{a} Indicator of Omicron-BA.2 ($\alpha_4$) &  \pmb{4.158 (1.543)} \\
               \tnote{a} Indicator of Omicron-BA.3+ ($\alpha_5$) &  \pmb{4.592 (1.739)}  \\
              \midrule
             \tnote{b} $\rho$ during initial  ($\rho_0 = \frac{1}{1+e^{\lambda_0}}$) &  \pmb{0.454 (0.042)}  \\
               \tnote{b} $\rho$ during  Alpha  ($\rho_1 = \frac{1}{1+e^{\lambda_0+\lambda_1}}$) &  \pmb{0.086 (0.010)}  \\
               \tnote{b} $\rho$ during  Delta  ($\rho_2 = \frac{1}{1+e^{\lambda_0+\lambda_2}}$) &  \pmb{0.999 (0.003)}  \\
               \tnote{b} $\rho$ during  Omicron-BA.1  ($\rho_3 = \frac{1}{1+e^{\lambda_0+\lambda_3}}$) &  \pmb{0.096 (0.011)}  \\
               \tnote{b} $\rho$ during Omicron-BA.2  ($\rho_4 = \frac{1}{1+e^{\lambda_0+\lambda_4}}$) &  \pmb{0.999 (0.002)}  \\
              $\rho$ during Omicron-BA.3+  ($\rho_5 = \frac{1}{1+e^{\lambda_0+\lambda_5}}$) &  0.522 (0.065)  \\
              \bottomrule
		\end{tabular}
\begin{tablenotes}\footnotesize
\item [a] Bold estimates of regression coefficients $\alpha_1$ to $\alpha_5$ indicate the estimate is significantly different from 0, which implies the corresponding period has significant different intercept compare to initial period.
\item [b] Bold estimates of $\rho_1$ to $\rho_5$ indicate the estimate is significantly different from $\rho_0$, which implies the distribution of $L$ is significantly different from initial period. 
\end{tablenotes}
                \caption{Estimates of the parameters $(\alpha_0,\pmb{\alpha}, \pmb{\rho})$ for model considered in Example S1.}
\label{tableexample1.5}
\end{threeparttable}
\end{table}

\begin{table}[!htb]
\centering
\begin{threeparttable}
\centering
            \begin{tabular}{l r r}
			   \toprule
			   & \textbf{Omicron BA.1} & \textbf{Omicron BA.2+}  \\
                  \textbf{Parameter} & \textbf{Estimates ($SE$)}& \textbf{Estimates ($SE$)}   \\
			   \midrule
	 	  Intercept ($\alpha$) & -21.331 (1.520) &  9.632 (.988) \\
               Standard deviation  ($\sigma$) & 6.117 (.483) &   9.089 (.315) \\
            \midrule
             Intercept of $\rho(\tau;\pmb{\lambda})$  ($\lambda_0$) & 9.999 (2.164) &  9.898 (2.379)\\
             \tnote{a} Change rate of $\rho(\tau;\pmb{\lambda})$ ($\lambda_1$) & \pmb{-0.382 (.091)} & \pmb{-0.391 (.095)} \\
                 & & ``hard" stratification  \\
                 \midrule
                 \midrule
	 	  Intercept ($\alpha$) & 4.782 (.727) &  6.295 (.811) \\
               Standard deviation  ($\sigma$) & 5.482 (.435) &  7.225 (.524) \\
            \midrule
             Intercept of $\rho(\tau;\pmb{\lambda})$  ($\lambda_0$) & 2.618 (.089) &  9.377 (2.991)\\
             \tnote{a} Change rate of $\rho(\tau;\pmb{\lambda})$ ($\lambda_1$) & \pmb{-0.159 (.019)} & \pmb{-0.365 (.094)} \\
                              & & ``soft" stratification  \\

              \bottomrule
		\end{tabular}
\begin{tablenotes}\footnotesize
\item [a] Bold estimates of $\lambda_1$ indicate significance deviating from 0, which indicate the importance of the $\tau$-varying transition probability.
\end{tablenotes}

                \caption{Estimates of the parameters $(\alpha, \sigma, \pmb{\lambda})$ for stratified analysis in Section 2.2 in the supplementary file. The Omicron BA.1 period in Ottawa is from 2021 Nov 20th to 2022 March 21st; the Omicron BA.2+ period in Ottawa is from 2022 March 21st to 2022 Nov 13rd. The stratification is a one-to-one correspondence with the stratification in Figure \ref{hardstratified2} for ``hard" stratification, and Figure \ref{softstratified2} for ``soft" stratification.}
\label{tablestratified2}
\end{threeparttable}

\end{table}